\documentclass[useAMS,usenatbib]{aa}

\usepackage{graphicx}
\usepackage{natbib}
\usepackage{longtable}
\usepackage{multirow}
\usepackage[varg]{txfonts}
\usepackage{amsmath}
\usepackage{bm}
\usepackage{txfonts}
\begin{document}

\title{DustPedia - the relationships between stars, gas and dust for galaxies residing in 
different environments}
\titlerunning{DustPedia}
\author{J. I.~Davies$^{1}$, 
A.~Nersesian$^{2,4}$,
M.~Baes$^{2}$, 
S. Bianchi$^{3}$, 
V. Casasola$^{3}$, 
L.~P. Cassar\`a$^{4}$, 
C. J. R.~Clark$^{1}$,
I.~De~Looze$^{5}$, 
P.~De~Vis$^{6}$,
R.~Evans$^{1}$,
J.~Fritz$^{7}$,
M.~Galametz$^{8}$, 
F.~Galliano$^{8}$, 
A.~P.~Jones$^{6}$,      
S.~Lianou$^{8}$, 
S.~C.~Madden$^{8}$, 
A.~V.~Mosenkov$^{9}$, 
M.~W.~L.~Smith$^{1}$, 
S. Verstocken$^{2}$, 
S.~Viaene$^{2,10}$,
M.~Vika$^{4}$,
E. Xilouris$^{4}$, 
N.~Ysard$^{6}$\\ 
}
\institute{
$^{1}$ School of Physics and Astronomy, Cardiff University, The Parade, Cardiff CF24 3AA, UK.  \\
\email{Jonathan.Davies@astro.cf.ac.uk}  \\
$^{2}$ Sterrenkundig Observatorium, Department of Physics and Astronomy, Universiteit Gent Krijgslaan 281 S9, B-9000 Gent, Belgium. \\
$^{3}$ INAF-Osservatorio Astrofisico di Arcetri, Largo E. Fermi 5, I-50125, Florence, Italy. \\
$^{4}$ National Observatory of Athens, Institute for Astronomy, Astrophysics, Space Applications and Remote Sensing, Ioannou Metaxa and Vasileos Pavlou GR-15236, Athens, Greece. \\
$^{5}$ Department of Physics and Astronomy, University College London, Gower Street, London WC1E 6BT, UK. \\
$^{6}$ Institut d'Astrophysique Spatiale, CNRS, Univ. Paris-Sud, Universit\'e Paris-Saclay, B\^{a}t. 121, 91405, Orsay Cedex, France. \\
$^{7}$ Instituto de Radioastronom\'\i a y Astrof\'\i sica, UNAM, Campus Morelia, A.P. 3-72, C.P. 58089, Mexico. \\
$^{8}$ Laboratoire AIM, CEA/DSM - CNRS - Universit\'e Paris Diderot, IRFU/Service d'Astrophysique, CEA Saclay, 91191, Gif-sur- Yvette, France. \\
$^{9}$ Central Astronomical Observatory of RAS, Pulkovskoye Chaussee 65/1, 196140, St. Petersburg, Russia. \\
$^{10}$ Centre for Astrophysics Research, University of Hertfordshire, College Lane, Hatfield AL10 9AB, UK. \\
}
\date{Recieved; ?}

\titlerunning{DustPedia}
\authorrunning{Davies et al.}



\abstract
{We use a sub-set of the DustPedia galaxy sample (461 galaxies) to investigate the effect the environment has had on galaxies. We consider Virgo cluster and field samples and also assign a density contrast parameter to each galaxy, as defined by the local density of SDSS galaxies. We consider their chemical evolution (using $M_{Dust}/M_{Baryon}$ and $M_{Gas}/M_{Baryon}$), their specific star formation rate ($SFR/M_{Stars}$), star formation efficiency ($SFR/M_{Gas}$), stars-to-dust mass ratio ($M_{Stars}/M_{Dust}$), gas-to-dust mass ratio ($M_{Gas}/M_{Dust}$) and the relationship between star formation rate per unit mass of dust and dust temperature ($SFR/M_{Dust}$ and T$_{Dust}$). Late type galaxies (later than Sc) in all of the environments can be modelled using simple closed box chemical evolution and a simple star formation history ($SFR(t) \propto t\exp{-t/\tau}$). For earlier type galaxies the physical mechanisms that give rise to their properties are clearly much more varied and require a more complicated model (mergers, gas in or outflow). However, we find little or no difference in the properties of galaxies of the same morphological type within the cluster, field or with different density contrasts. It appears that it is morphology, how and whenever this is laid down, and consistent internal physical processes that primarily determine the derived properties of galaxies in the DustPedia sample and not processes related to differences in the local environment.}


\maketitle

\section{Introduction} 
The hierarchical structure formation model  predicts that dark matter from the primordial density field collapses into virialised haloes, which then provide the gravitational potential wells for the infall of baryons and subsequently the formation of galaxies (White and Rees 1978). 

As galaxies age they are expected to interact with their environment through their ability to grow by utilising infalling gas and from the consumption of other galaxies that merge with them (Springel et al. 2005). Thus, according to this model, just what a galaxy is today should to some extent be a reflection of the environment it continually finds itself within, which provides resources for its future development\footnote{For the purposes of this paper "environment" will be defined by the local number density of galaxies.}. In support of this scenario numerical simulations indicate that the environment can have quite a large influence on the physical properties of galaxies (De Lucia et al. 2012, Yang et al. 2018). 

Probably the most clearly defined relationship between environment and a galaxy's physical properties is that related to its morphology (E, Sa, Sb etc.). Oemler (1974) reported the increased fraction of early type galaxies in clusters. This was then further described by Dressler (1980), who showed how the morphological mix of galaxies in clusters relates to the local galaxy density - higher densities leading to a larger fraction of early types. Both of these works relied on observations and comparisons of galaxies in clusters compared to those in the general field.

Since then much more evidence has accumulated to firmly establish that many global physical parameters of galaxy populations, for example,  stellar mass, galaxy colour, star formation rate and gas content also strongly depend on environment (Ostriker and Tremaine 1975, Dressler 1980, Giovanelli and Haynes 1985, Kodama et al. 2001, Gavazzi et al. 2002, Kauffman et al. 2004, Casasola et al., 2004, Baldry et al. 2006, Fumagalli et al. 2009, Peng et al. 2010). However, understanding whether this is an ongoing environmental effect as galaxies evolve is confused by the galaxy morphology density relation. In denser environments (for example clusters) there are many more quiescent 'early' type galaxies, which almost by definition have lower star formation rates and less gas (Dressler et al. 1985). 

The galaxy population is clearly evolving with there being many less massive galaxies at high redshift and an apparent lack of the Hubble sequence beyond a redshift of 2 (Conselice et al. 2005, Buitrago et al. 2013). However, it is still not known whether morphology once obtained is fixed over a long period of time or whether morphological transformation occurs, and if so how many changes in morphology may occur (Conselice et al. (2014). van der Kruit and Freeman (2011) point out that galaxies of the same luminosity (and many other similarities, they compare the LMC with M33)  can have very different morphologies and use this to support the idea of morphological transformation. 

A major difficulty in trying to understand this issue is that there is no definitive model of how morphology itself arises (Boselli and Gavazzi 2006, Wel et al., 2010) or why the morphology density relation exists (see discussion in Weinmann et al. 2006). It has been plausibly suggested that morphology may have its origin in some combination of merger history and angular momentum (Rodriguez-Gomez et al. 2016, Cortese et al. 2016a), mergers being invoked as the explanation of both spheroidal structure and morphological transformation between different types. However, recent observations and numerical simulations have shown that spheroidal structures can grow in isolation as stars form in-situ removing the necessity of mergers (Sales et al. 2012, Lofthouse et al. 2017, Rodriguez-Gomez et al. 2017). 

 Morphology does appear to be linked to other galaxy properties. For example more massive galaxies tend to be more spheroidal and star forming galaxies  tend to have more prominant discs (Gadotti 2009, Whitaker et al. 2015). Probably related to this is the high abundance of generally spheroidal red and/or gas deficient galaxies in higher density environments, particularly galaxy clusters (Visvanathan and Sandage, 1977, Giovanelli and Haynes 1985) and the bimodality of the star formation rate (SFR). This bimodality leads to a clear distinction between galaxies that are actively star forming and those that are "quenched" and their corresponding disc or spheroidal structure (Strateva et al. 2001, Blanton et al. 2005). These two populations are clearly evolving as the relative numbers of quenched galaxies has increased since a redshift of one (Tomczak et al. 2014).
 
With regard to the SFR of cluster galaxies compared to the field, the situation is a little less clear. There is evidence for the suppression of star formation in clusters when compared to the field, even when the morphological mix is taken into account (Kennicutt 1983, Dressler et al. 1985, Bamford et al. 2009). On the other hand the interaction of late type galaxies with the cluster environment may lead to enhanced star formation (Kennicutt et al. 1984, Moss and Whittle 1993). Willett et al. (2015) found no difference in the SFR stellar mass relation for star forming galaxies of different morphological types suggesting that galaxies are strongly self-regulated (see also Bait et al. 2017). 

Although the cluster morphology density relation is clearly well defined (Dressler 1980), the situation is far less clear outside of galaxy clusters (Kauffmann et al. 2004). Although there is good evidence that galaxy properties, such as the mean SFR, depend on the local galaxy density (Gomez et al., 2003, Malavasi et al., 2017, Kuutma et al., 2017), it is not clear how much of this depends on the local morphological mix. So, again this begs the question whether it is  the environment affecting the mean star formation rate of galaxies or whether it is the result of the predetermined morphological mix of the galaxies. Thus, we might question whether there  really is a strong on-going environmental effect on galaxy physical parameters or whether what is observed is more closely related to some pre-defined morphological mix of the galaxies. 

Many environmentally dependent mechanisms for potentially altering the properties of galaxies have been proposed. As well as a complex merger history (Conselice et al. 2003, Smethurst et al. 2015), these also include interactions that remove or restrict the supply of gas, such as ram pressure stripping, 'strangulation' and 'starvation' (Poggianti et al. 2017, Weinmann et al. 2009). There are also gravitational processes that disrupt and possibly strip stars and gas from galaxies and more subtle gravitational interactions like harassment (Moore et al. 1996). All of these are expected  to be ongoing as a galaxy evolves, but it is not clear how influential they might be or whether they are actually secondary effects, with the real defining parameters being what happens at the very start, when a galaxy begins to form.  

In support of this latter suggestion there are other observations that have indicated that the environment is having very little effect on the observed properties of galaxies.
For example, Park et al. (2007) essentially say that all of the above possible environmental effects are  ineffective. From a large sample of SDSS galaxies they conclude that if a galaxy's luminosity and morphology are fixed then all other properties such as colour, colour gradient, concentration, size, velocity dispersion, and SFR are essentially independent of environment. The important implication of this is that luminosity and morphology are pretty much fixed at the start, and then all else follows on. The luminosity may play an important role as Robotham et al. (2014) suggest that the characteristic absolute magnitude (M$^{*} \approx -20.0$ in the $g$ band, $\simeq 10^{10}$ $M_{\odot}$) may be the dividing line between those galaxies that grow via in-situ star formation (M$^{*}>-20.0$) and those that have predominately grown through mergers (M$^{*}<-20.0$).

In this paper we particularly consider a well observed sample of local (within 3000 km s$^{-1}$) galaxies taken from the DustPedia database (Davies et al. 2017). These are all nearby galaxies that have been observed as part of various Herschel Space Observatory programmes and so have many imaging observations across the spectral energy distribution (SED) from the ultra-violet to the far-infrared. The DustPedia sample probes a range of baryonic mass densities - containing galaxies that reside in the "field" and in the $\approx100$ times denser Virgo cluster (Davies et al. 2014) thus enabling us to consider environmental influences\footnote{We note that with our data we are only able to compare the field with one cluster (Virgo), which is not as rich or evolved as some clusters and so the environmental effects we are looking to identify may not be as obvious when compared to other richer clusters.}.

In what follows we will first of all define the galaxy sample to be used, which has its origins in legacy data from the Herschel Space Observatory data archive (section 2).  Then we will describe how we calculate SFRs, stellar, gas and dust masses, and dust temperatures (section 3). In section 4 we will consider environmental effects firstly as a function of whether the galaxy belongs to the Virgo cluster or not and secondly via a calculated local galaxy density. Finally we will look at how, if at all chemical evolution (using $M_{Dust}/M_{Baryon}$ and $M_{Gas}/M_{Baryon}$), specific star formation rate ($SFR/M_{Stars}$), star formation efficiency ($SFR/M_{Gas}$), stars to dust mass ratio ($M_{Stars}/M_{Dust}$), gas to dust mass ratio ($M_{Gas}/M_{Dust}$) and the relationship between star formation rate per unit mass of dust and dust temperature ($SFR/M_{Dust}$ and T$_{Dust}$) have been affected by the local environment. Where necessary we will use a cosmological model where $H_{0}=67.8$ km s$^{-1}$ Mpc$^{-1}$ and $\Omega_{M}=0.31$ (Ade et al., 2016).

\section{The data}
The data used here is taken from the DustPedia database\footnote{http://dustpedia.astro.noa.gr/}. DustPedia is an European Union funded project\footnote{DustPedia is a collaborative focused research project supported by European Union Grant 606847 awarded under the FP7 call. Further information can be found at www.dustpedia.com.} to exploit the legacy value of the Herschel Science Archive (HSA)\footnote{http://archives.esac.esa.int/hsa/aio/doc/}. Full details of the DustPedia project i.e. the data used, the models developed and the science objectives, can be found in Davies et al. (2017). Additional details of the ancillary data used, the photometry and flux extraction methods can be found in Clark et al. (2018). 
  
 In summary the DustPedia sample consists of all galaxies observed by Herschel that lie at recessional velocities of
$< 3000$ km s$^{-1}$, with optical diameters $> 1$ arc min and a WISE\footnote{Wide-Field Infrared Survey Explorer, https://www.nasa.gov/mission\_pages/WISE/ } 3.4$\mu$m signal-to-noise ratio $> 5$. The velocity restriction means that we include galaxies that are "local" yet still reside in different environments. For example this selection includes the Virgo and Fornax clusters and galaxies in the super-galactic plane that consists of the Virgo southern extension and connecting galaxies from Virgo to the less rich Ursa Major cluster (see Fig. 1 in Davies et al. 2017). The total sample consists of 875 galaxies excluding the four very large angular sized galaxies M31, M33, LMC and SMC. The sample contains galaxies with a wide range of morphological types (T) with a rather flat distribution of galaxy numbers across the range of T types (see Table 3 and Fig. 2 in Davies et al. 2017). 
  
 Herschel imaging data (70 - 500 $\mu$m) obtained using PACS (Poglitsch et al., 2010) and SPIRE (Griffin et al., 2010) have been extracted from the Herschel Science Archive and reduced in a uniform manner and calibrated using the HIPE v13 software package\footnote{HIPEv13 was the then current release of the Herschel Interactive Processing Environment (Ott, 2010): http://www.cosmos.esa.int/web/herschel/hipe-download.} (Davies et al. 2017, Clark et al. 2018). This provides data at typically five points across the far infrared SED (70 to 500$\mu$m). 
 
 Along with the far infrared data we have assembled a large amount of ancillary data ranging from the ultra-violet (GALEX)\footnote{Galaxy Evolution Explorer, https://asd.gsfc.nasa.gov/archive/galex/}, through the optical (SDSS)\footnote{Sloane Digital Sky Survey, http://www.sdss.org/} to the near (2MASS)\footnote{2 Micron All Sky Survey,  http://www.ipac.caltech.edu/2mass/} and mid-infrared (WISE)\footnote{The Wide-field Infrared Survey Explorer, http://wise2.ipac.caltech.edu/docs/release/allsky/}. For some of the brighter galaxies the data is extended into the sub-mm/mm range (Planck)\footnote{https://www.cosmos.esa.int/web/planck.}. In total the DustPedia database is able to provide photometry in up to 41 bands.
 
Using the database images, we have performed aperture-matched multi-wavelength photometry for all sources. This suit of routines is a development of the CAAPR photometry pipeline used in Clark et al. (2015) and De Vis et al. (2017a), it is described in full in Clark et al. (2018). The full photometric data in all available bands can be obtained from the DustPedia database. 

Where galaxy distances are required we have used wherever possible velocity independent measures taken firstly from the Hyperleda\footnote{http://leda.univ-lyon1.fr/} database or if not available from there then from NED\footnote{http://ned.ipac.caltech.edu/}. If neither HyperLEDA nor NED redshift-independent distances are available for a source, we use the flow-corrected redshift-derived values provided by NED; these distances are calculated assuming a Hubble constant of H$_{0}$ = 73.24 km s$^{-1}$ Mpc$^{-1}$ (Riess et al. 2016, Clark et al., 2018).

\begin{table}
\begin{center}
\begin{tabular}{c|c|c|c|c}
Morphology & T-type & N & N$_{Field}$ & N$_{Virgo}$  \\ \hline
E/S0  & $T < 0.0$ & 113 & 65 & 48 \\
Sa/Sb  & $0.0 \le T \le3.0$ & 83 & 53 & 30  \\
Sc/Sd  & $3.0 < T \le 6.0$ & 155 &124 & 31  \\
Sm/Irr  & $T >6.0$ & 110 & 76 & 34  \\
\end{tabular}
\caption{The numbers of different morphological types of Dustpedia galaxies that fall within the SDSS footprint and have sufficient information to derive, stellar mass, gas mass, SFR, SF history, dust mass and temperature. N is the number of galaxies of each morphological type (N$_{Field}$, N$_{Virgo}$ respectively for those inside and outside of the Virgo cluster).}
\label{table:numbers_morph}
\end{center}
\end{table}  

As a means of quantifying the local environment of DustPedia galaxies (see section 4.2 below) we have also obtained additional data from the SDSS. To this end in this paper we only consider a sub-set of DustPedia galaxies that overlaps with the SDSS footprint ($120.0 \le$RA(J2000)$\le 240.0$, $0.0 \le$Dec(J2000)$\le 60.0$). In addition we also require that the SED of each galaxy can be successfully fitted by the CIGALE SED fitting package and that an atomic hydrogen mass is also available (see below). This reduces the DustPedia sample used here to a sub-sample of 461 galaxies, which can then also be split into sub-samples of different morphological types (see Table \ref{table:numbers_morph}). 

There are 4618 SDSS galaxies with redshifts contained within approximately the same volume as the DustPedia sub-set. We have selected these to have redshifts between $z=-0.003$ ($v$=-1000 km s$^{-1}$) and $z=0.0117$ ($v$=3500 km s$^{-1}$) - some galaxies in our sample have negative velocities and 3500 km s$^{-1}$ is used as some DustPedia galaxies may have companions at slightly higher velocities than the DustPedia survey limit of 3000 km s$^{-1}$. The median $g$ band apparent and absolute magnitude of the DustPedia galaxies are 12.7 and -18.9 respectively.  This compares with the median $g$ band apparent and absolute magnitudes of the SDSS galaxies of 16.6 and -15.6 respectively, enabling us to identify fainter companions to our DustPedia sample galaxies. How we use the SDSS data in conjunction with the DustPedia data is explained in section 4.2 below.

\begin{figure*}
\centering
\vspace{-0.0cm}
\includegraphics[scale=0.55]{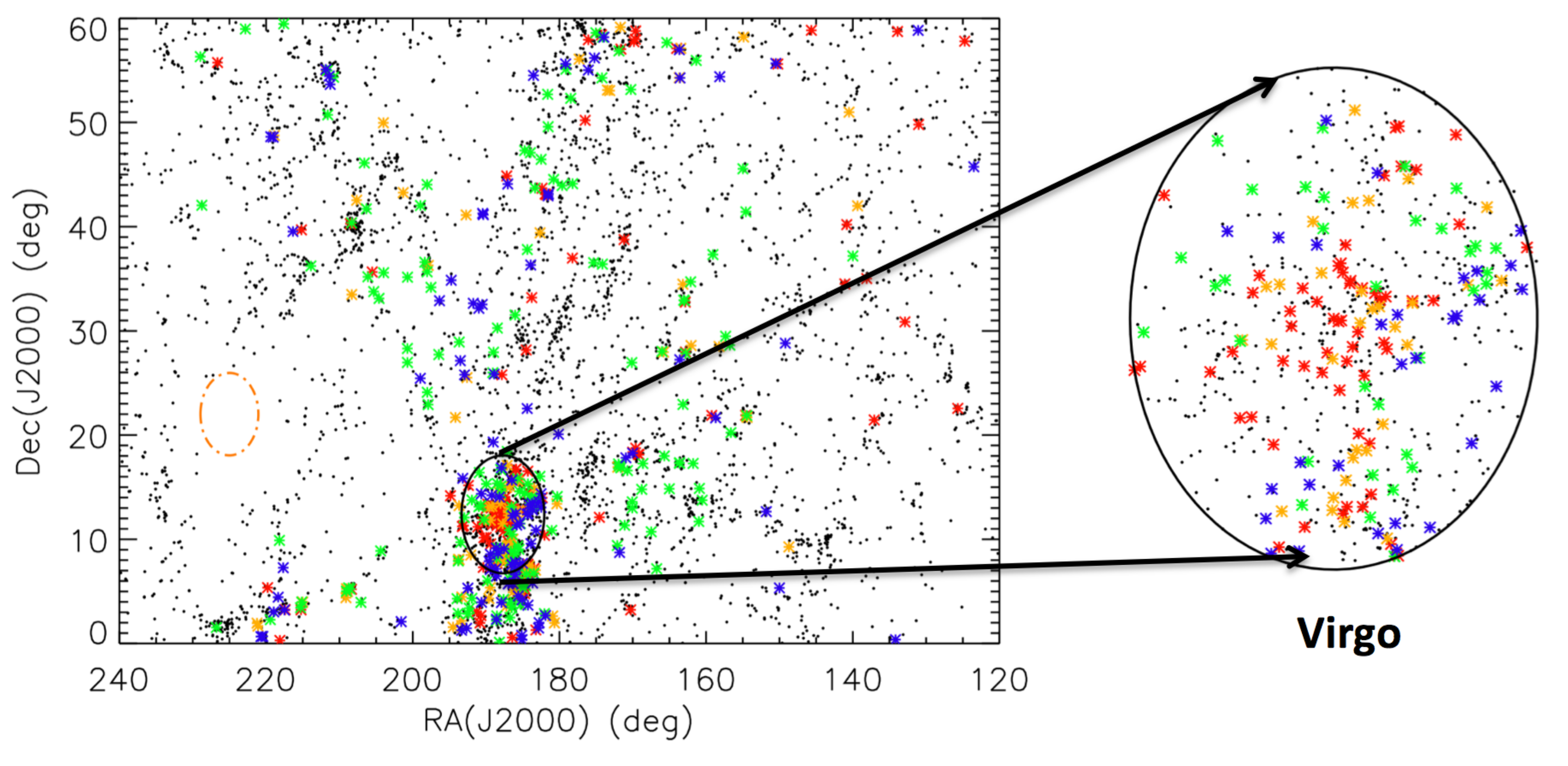}
\vspace{-0.0cm}
\caption{The spatial distribution of galaxies over the area of sky considered here. Black dots mark the position of 4618 galaxies selected from SDSS. Coloured stars are the DustPedia sample galaxies - E/S0 (red), Sa/Sb (yellow), Sc/Sd (green) and Sm/Irr (blue). The black ellipse marks the virial radius of the Virgo cluster centred on M87. The orange ellipse is the size of the density contrast aperture (radius of 1.5 Mpc, see section 4.2) at the sample median distance of 19.9 Mpc. On the right there is an expanded view of the Virgo cluster.}
\label{fig:fig_1}
\end{figure*}

Fig. \ref{fig:fig_1} shows both the spatial distribution of the DustPedia (split by morphological type) and SDSS sub-samples. Clearly the Virgo cluster is a prominent feature in our data and has been marked by the ellipse (virial radius of 5.65 degrees, McLaughlin, 1999) shown on Fig. \ref{fig:fig_1}. As described below we will also consider a Virgo cluster sample selected from within the cluster virial radius (143 galaxies) compared with a field sample consisting of those outside the cluster virial radius (318 galaxies).

Many of the conclusions and relations we can infer from our data could be biased by the rather ill-defined selection criteria we have used - the most significant (unquantifiable) criteria being that they were all observed as part of the many and varied projects carried out by the Herschel Space Observatory. KS tests comparing the distributions of our derived parameters, $M_{Star}$, $M_{Dust}$, SFR and $T_{Dust}$, for Virgo and the field, indicate that they are drawn from either different underlying distributions or that the selection criteria have biased the samples in different ways. It is impossible to decide on real differences or observational selection. As such here our intention is not to compare distributions of galaxy properties, i.e. relative numbers of a given mass, temperature or SFR, but to assume that each sample as a whole, Virgo and field, is "representative" of the underlying population, our sample being a random representation of the nature of the galaxies in the two environments.

\section{Parameters derived from the data}
We have used the results from the SED fitting package CIGALE (Boquien et al. 2019) to derive stellar mass ($M_{Star}$), dust mass ($M_{Dust}$), dust temperature ($T_{Dust}$) and SFR for each of our sample galaxies. A full description of how we have initiated, run, tested and compared the output from CIGALE is given in Nersesian et al. (2018). 

Very briefly, CIGALE is a software package that allows the user to create galaxy SEDs 
while taking into account the balance between the energy absorbed by dust in the UV-optical and then re-emitted 
in the infrared (Roehlly et al. 2014, Boquien et al. 2019). As part of the SED fitting we assume a SF history (Ciesla et al. 2016) along with the stellar emission from the stellar population models of Bruzual and Charlot (2003), a Salpeter initial mass function and solar metallicity. The dust extinction law used is a modified version of the standard starburst-like dust attenuation described by Calzetti et al. (2000), while for the infrared dust emission we have used our own (produced as part of the DustPedia project) dust grain model, called THEMIS (Jones et al. 2017)\footnote{see also: http://www.ias.u-psud.fr/themis/}. Using a Bayesian analysis CIGALE fits the available photometric data for each galaxy to derive $M_{Star}$, $M_{Dust}$ and a SFR. $T_{Dust}$ is derived from the mean intensity of the star light (Nersesian et al., 2018). AGN have not been excluded from our data, their possible and mainly small influence on the parameters we derive is discussed in Bianchi et al. (2018) and will only affect 2-3\% of the galaxies in the DustPedia sample.

We have also used the compilation of atomic gas masses from De Vis et al. (2018) to obtain values for the total gas mass of our sample galaxies. To convert from atomic mass to total gas mass (including molecular hydrogen and helium) we follow De Vis et al. (2018), who use the relation: \\
\begin{center}
$M_{Gas}=1.32 M_{HI}(1.0+0.17(M_{HI}/M_{Star})^{-0.72})$ 
\end{center}

\section{Galaxy properties within different environments}
We will firstly compare the measured properties of those galaxies that reside inside "Virgo" with those outside of the cluster virial radius (the "field"). Secondly, we will do a similar analysis, but for galaxies that reside in different environments as defined by the local density of SDSS galaxies.

\begin{figure}
\centering
\includegraphics[scale=0.25]{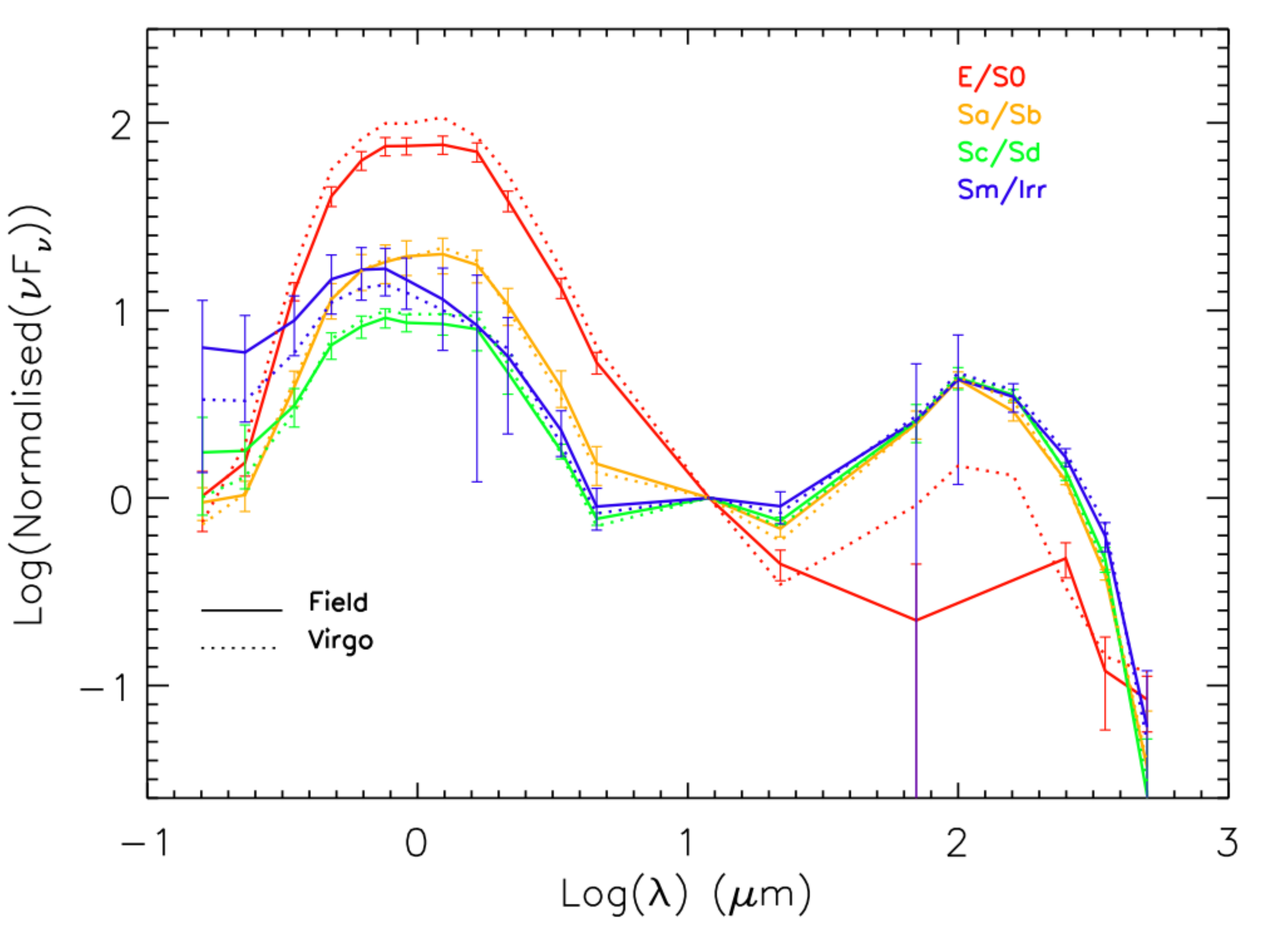}
\vspace{0cm}
\caption{The median spectra of galaxies as a function of their morphological type and whether they lie within the Virgo cluster or within the field. Error bars are approximately the same for the Virgo and field samples, but they are shown only for the latter as the plot otherwise becomes rather cluttered and confused.}
\label{fig:morph_spectra}
\end{figure}

\subsection{A comparison between those galaxies in the Virgo cluster and those in the field}
\subsubsection{Spectral energy distributions}
In Fig. \ref{fig:morph_spectra} we compare the mean SEDs of galaxies of different morphological types (normalised at 12$\mu$m). As expected there is a marked difference between the optical and far infrared outputs of the generally low dust mass early types (E/S0) and the later types. The latest types (Sm/Irr) have an up-turn in their emission in the blue/UV region again as expected given that they are generally active star formers, but also indicating that a significant fraction of their ultra-violet radiation escapes the dust. However, what is most interesting is the close similarity of the spectra for galaxies of the same morphology that reside within and outside of the Virgo cluster. In most cases the data for the two environments sit virtually on top of each other and there is no evidence that the SEDs of galaxies of the same morphology have been altered by the local environment. There are two small exceptions to this that we highlight, even though within the derived errors on the SEDs they are not strongly significant. 

Firstly, early (E/S0) field galaxies seem to produce less FIR radiation than those in the cluster even though their optical outputs are pretty much the same - it is not clear why this might be so. We will show below that cluster early types also have marginally hotter dust than those in the field, a plausible additional heating source for the cluster early types may come from the inter-galactic x-ray gas (Lebouteiller et al. 2017). Secondly, late type field galaxies (Sa to Irr, but significantly for Sm/Irr) have a blue/UV output that is higher than those in the cluster, even though their far infrared outputs are all about the same \footnote{Note that the error bars on individual points for this UV/blue "excess" are quite large, but it is a consistent trend over 5-6 data points.}. A parameter often used to measure extinction is the ultra-violet to total far infrared ratio (Gordon et al., 2000, Buat et al., 2007). Using this measure Fig. \ref{fig:morph_spectra} indicates that later types in the field suffer less extinction than Virgo galaxies of the same morphology. Whether this is because of differences in the spatial distribution of stars and dust or in the physical properties of the grains is not clear from the data we have here. 

\subsubsection{Chemical evolution}
One of the most fundamental models of how a galaxy may evolve concerns its chemical evolution i.e. given that a galaxy starts its life as a cloud of gas and finishes it as a collection of stellar remnants, what should we observe at various times as this process proceeds? 

\begin{figure}
\centering
\includegraphics[scale=0.32]{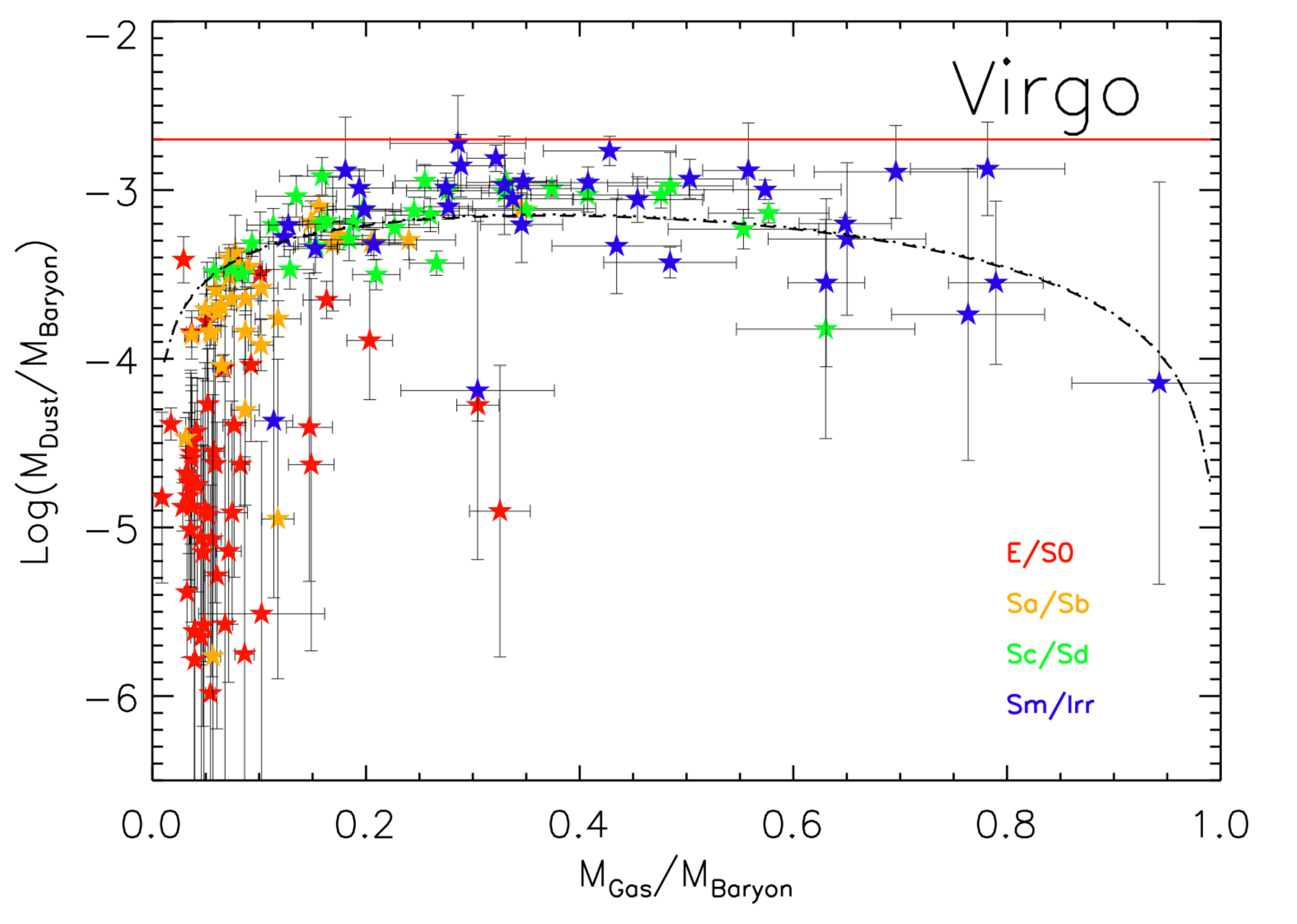}
\includegraphics[scale=0.32]{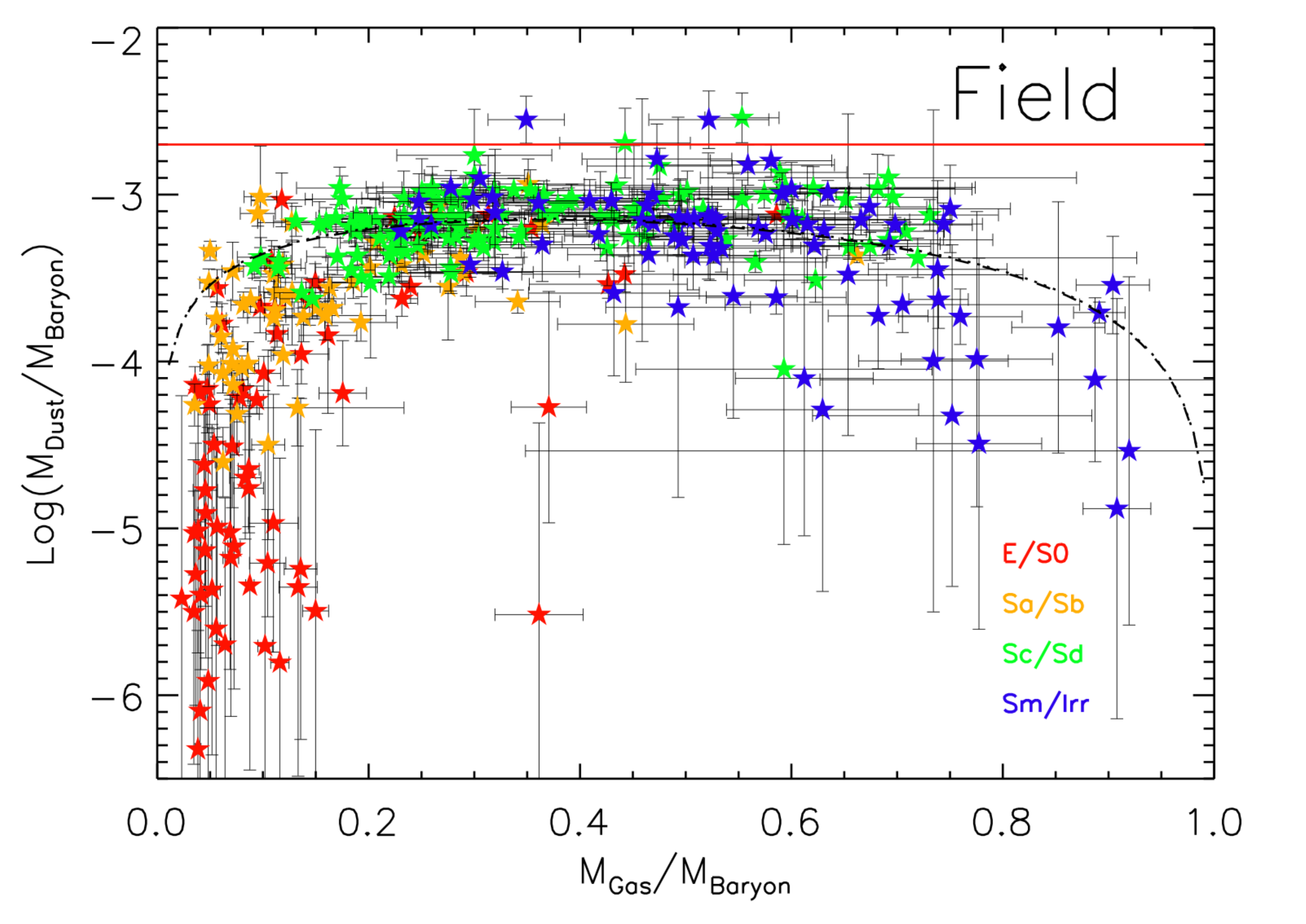}
\vspace{-0.2cm}
\caption{The relationship between dust and gas mass fraction as both a function of morphological type and environment - Virgo (top) and field (bottom). The dotted and dashed lines (they lie on top of each other and so are difficult to distinguish separately) correspond to a simple closed box chemical evolution model - Virgo dotted, field dashed line. The red line is the predicted maximum fractional dust mass for a dust-to-metals ratio of $\eta=0.5$ and a stellar yield of $p=0.01$.}
\label{fig:chem_evo}
\end{figure}

Within the bounds of a simple closed box chemical evolution model, Edmunds and Eales (1998) show that the fractional maximum dust mass ($\Delta_{max,f}$) at a given gas fraction ($f$) is given by:\\
$\Delta_{max,f}=\eta pf \ln{(1/f)}$ \\
where $\eta$ is the fraction of the interstellar metals in dust and $p$ is the stellar yield of heavy elements, $f$ is the fraction of the baryonic mass that remains in the gas. The model prediction is illustrated on Fig. \ref{fig:chem_evo} by the black line. This is a single unified model and so does not distinguish between different evolutionary paths for galaxies of different morphologies. By comparing galaxies of different morphologies we are assuming that they are linked in their evolution only through changes in their gas fractions and so, for example, galaxies with low gas fractions were once the same as galaxies at a higher gas fraction in an earlier life. 

Looking at Fig. \ref{fig:chem_evo} this seems to be a reasonable  interpretation of the data. There is a systematic  progression from high to low gas fractions as morphology changes from late to early along with predictable changes in the fraction of baryonic mass in the dust. In this context we can interpret the large range in dust mass fraction for early types as being a consequence of the predicted steep fall in dust mass as a galaxy exhausts its gas. This simple model also provides us with a normalisation and hence a consideration of the product $\eta p$ (see below). However, we note that in detail the Edmunds and Eales (1998) model does not provide a good fit at the earliest and latest stages of evolution. For example De Vis et al. (2018) show that at gas fractions below 0.15, the model overestimates the dust-to-baryon ratio by an order of magnitude. More detailed models with gas outflows and dust growth in the inter-stellar medium are explored in De Vis et al. (2018).

Considering this simple chemical evolution model we might hypothesise that the known gas depletion of Virgo galaxies, when compared to galaxies in the field of the same morphological type (Giovanelli and Haynes, 1985), is not due to gas stripping, but just that they are further along their evolutionary path.  The measured gas depletion for late type galaxies (Sc/Sd, Sm, Irr) in the cluster compared to the field - median values of $M_{Gas}/M_{Baryon}$ are $0.43\pm0.02$ and $0.28\pm0.03$ for field and cluster respectively - shows that the same depletion effect is present in our sample. 

To see if there are differences in the chemical evolution of  galaxies in the cluster and field, as there might be if, for example cluster galaxies preferentially lose gas, we have fitted the maximum dust mass function ($\Delta_{max,f}$) to the data. This is only a normalisation, as the shape of the curve is fixed. Using a non-linear least squares fit to the function we find almost identical normalisations for both the field and cluster (black dotted and dashed lines, Fig. \ref{fig:chem_evo}). 

The above normalisation gives a value for the product $\eta p$. We obtain values of $\eta p=2.0\pm0.1\times10^{-3}$ and $1.9\pm0.1\times10^{-3}$ for field and cluster respectively. Edmunds and Eales (1998) originally gave values of $\eta=0.5$ and $p=0.01$, which leads to the maximum value of $M_{Dust}/M_{Baryon}$ indicated by the red line on Fig. \ref{fig:chem_evo} and a value of $\eta p=5.0 \times 10^{-3}$, somewhat larger than the value we measure. Recently De Vis et al. (2018) have used the DustPedia sample to measure a smaller value of $\eta=0.283$. This gives a value of $\eta p=2.83 \times 10^{-3}$ if $p=0.01$, closer to the value we measure. Using our value for the product ($\eta p=2.0\times10^{-3}$) and the De Vis (2018) value for $\eta=0.283$ leads to a value of $p=7.1\times10^{-3}$ consistent with the range of values measured by Davies et al. (2014) -  $p=3.0-12.0\times10^{-3}$ depending on morphological type. 

Edmunds (1990) define an effective yield ($p_{eff}=\frac{z}{\ln{1/f}}$, where $z$ is the metalicity), which is the derived yield irrespective of whether there are inflows or out flows of gas. This extends the usefulness of the chemical evolution model to situations other than just closed box evolution. That we measure an almost identical value of $\eta p$ for field and cluster galaxies indicates, unless there is some conspiracy between $\eta$ and $p$, that $p \equiv p_{eff}$ is not changing between field and cluster. $p_{eff}$ is expected to change if the chemical evolution is being affected by, for example gas stripping (lower values of $p_{eff}$) in the cluster environment. There is no evidence that the chemical evolution of galaxies in the cluster and field is any different, even though Virgo galaxies are currently depleted in gas compared to the field.

\subsubsection{The specific star formation rate}
\begin{figure}
\centering
\includegraphics[scale=0.25]{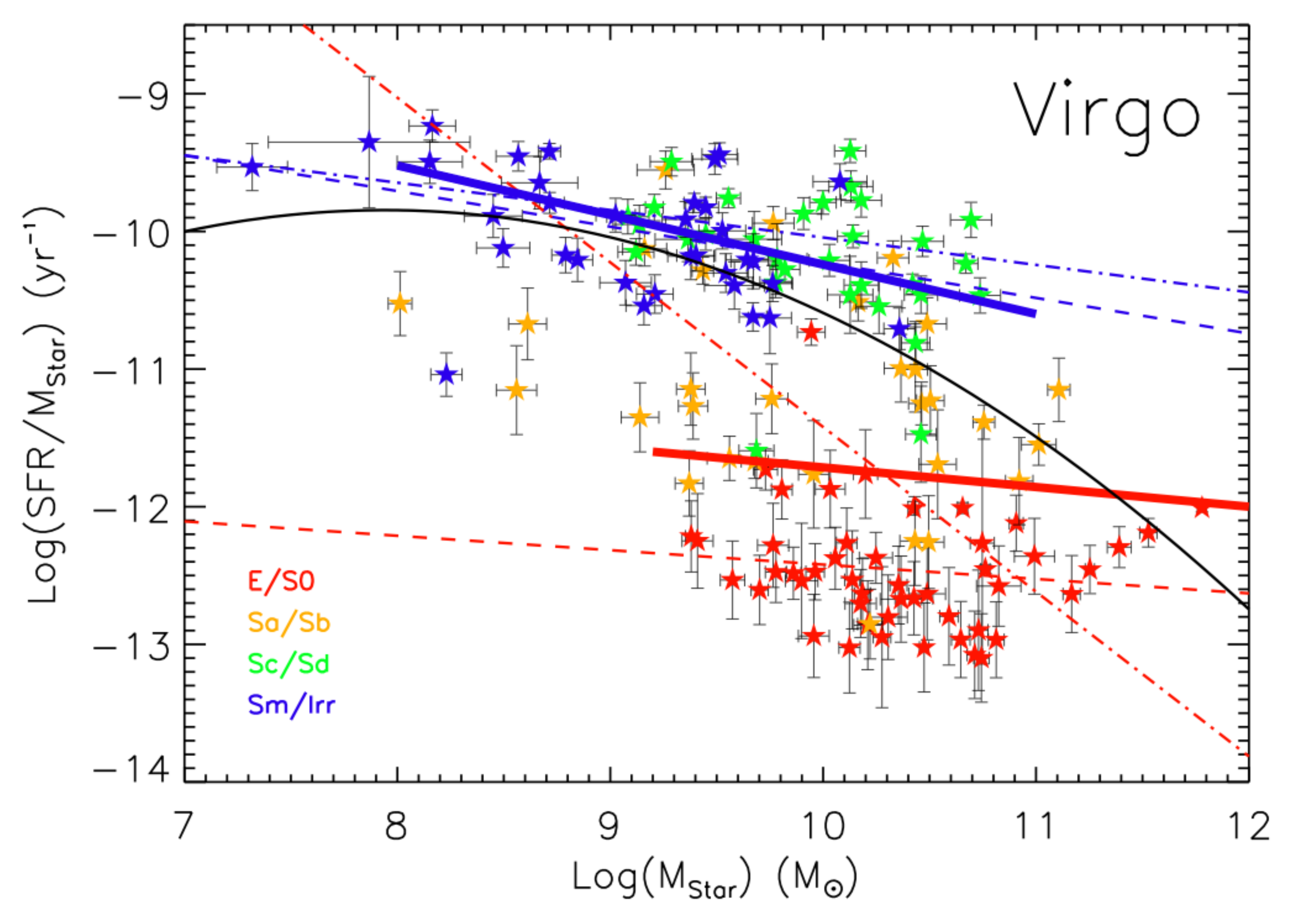}
\includegraphics[scale=0.25]{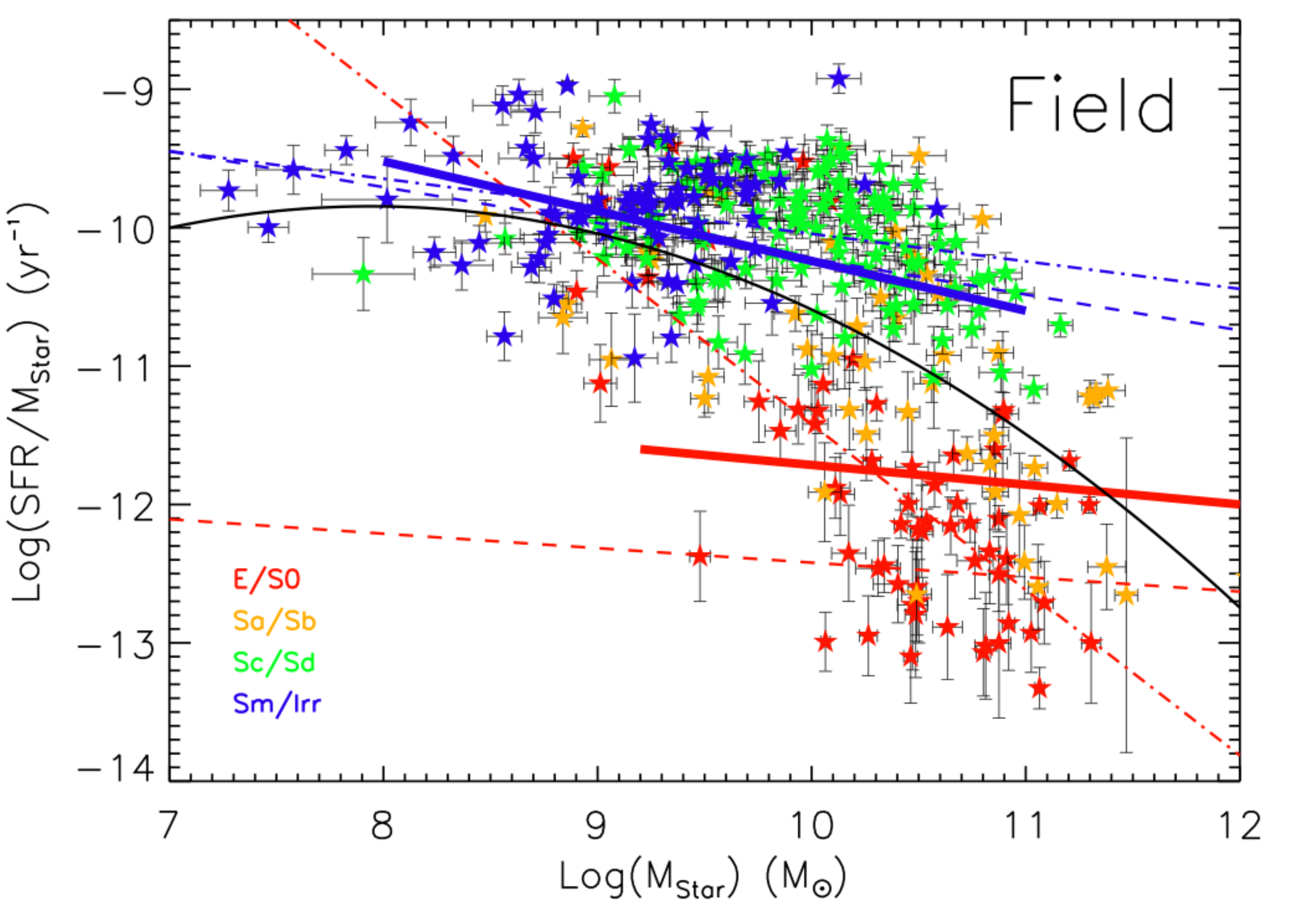}
\vspace{-0.2cm}
\caption{The 'main sequence' for our data defined by the stellar mass ($M_{\odot}$) and SSFR ($yr^{-1}$). Galaxies are distinguished by their morphological type and whether they are members of the Virgo cluster (top) or in the field (bottom). The thick blue line is the locus of the galaxy main sequence and the red line the locus of "non-star forming" galaxies as defined by Schiminovich et al. (2007). For comparison the dashed lines are linear fits to the late (Sc/Sd, Sm/Irr) and early types (E/S0) in Virgo (dash) and the field (dot-dash). The solid black line is the proposed fit to the data used by Eales et al., (2017).}
\label{fig:sfr_stellar_mass}
\end{figure}

A missing ingredient from the chemical evolution model is one that links a galaxy's properties to the way in which it is forming stars. In Fig. \ref{fig:sfr_stellar_mass} we have plotted the specific SFR (SSFR, i.e. SFR per unit stellar mass) against the stellar mass for our sample of galaxies. The "star forming" galaxies (types later than Sa) in both Virgo and the field fall predominately on what is often described as a "main sequence"  and clearly this is an almost identical line for both field and cluster galaxies (Fig. \ref{fig:sfr_stellar_mass}, blue dashed and dot-dashed lines respectively). Both Peng et al. (2010) and Calvi et al. (2018) have previously noted that SFR and SSFR as a function of stellar mass does not vary with environment. From Fig. \ref{fig:sfr_stellar_mass} it is also clear that our main sequence is almost identical to that defined by the previous work of Schiminovich et al. (2007), using a large sample of SDSS galaxies (as shown by the thick blue line on Fig. \ref{fig:sfr_stellar_mass}). 

Interestingly the early type galaxies (E/S0) in our Virgo sample also seem to reside on a well defined "(main) sequence" but with lower (two orders of magnitude) values of SSFR than that of the later types. The cluster in particular seems to have well defined star forming and quiescent sequences and a few predominantly Sa/Sb galaxies occupying the "green valley" between the two. It has previously been proposed that there is an evolutionary link between the two sequences due to the gradual consumption or loss of gas, which depresses star formation (Dunne et al. 2011, Genzel et al. 2015, Scoville et al. 2016). Most of the Virgo early types reside below the loci of "non-star forming galaxies" as defined by the thick red line (also from Schiminovich et al. 2007), but in their paper they caution the reader that unlike the star forming sequence, in this part of the diagram their line is best taken as an upper limit. This is because it remains extremely difficult to probe star formation at these levels. 

\begin{table}
\begin{center}
\begin{tabular}{c|c|c}
  Type & $-\log{(<SSFR>)}_{Field}$ & $-\log{(<SSFR>)}_{Virgo}$   \\
   &  ($yr$) & ($yr$) \\ \hline
E/S0 & $11.8\pm0.1$ & $12.5\pm0.1$ \\
Sa/Sb & $11.0\pm0.1$ & $11.2\pm0.1$ \\
Sc/Sd & $10.1\pm0.1$ & $10.2\pm0.1$ \\
Sm/Irr & $9.8\pm0.1$ & $10.0\pm0.1$ \\
\end{tabular}
\caption{The mean time to form the current mass of stars at the current SFR (SSFR$^{-1}$) for different morphological types both within the Virgo cluster and in the field.}
\label{table:specific_sfr}
\end{center}
\end{table} 

While the Virgo cluster data seems to conform to the idea of two sequences with evolution between them, as described above, this is not so clear-cut for the field data (Fig. \ref{fig:sfr_stellar_mass}, bottom). Eales et al. (2017) have previously suggested that Fig. \ref{fig:sfr_stellar_mass} actually shows a single sequence of morphological type, roughly from top left to bottom right as shown on Fig. \ref{fig:sfr_stellar_mass} by the solid black line (for comparison see Fig. 2 in Eales et al., 2017). There is clearly a change in morphology as one moves along the black line, but given the rather distinct sequences and simple interpretation of the cluster data (Fig. \ref{fig:sfr_stellar_mass}, top) we suggest that the star forming "main sequence" is a consistent feature of galaxies. Possibly there are small off-sets between the positions of the main sequence for galaxies formed in different places in the Universe that leads to more scatter in Fig. \ref{fig:sfr_stellar_mass} for the field galaxies, when compared to that in the cluster - this may be related to small differences in their star formation history (see below). 

The SSFR (Fig. \ref{fig:sfr_stellar_mass}) relates the current SFR to the resultant sum of star formation over cosmic time. As might be expected the position of the main sequence changes with cosmic time and so this is a measure of the growth of stellar mass (Daddi et al. 2007). As already noted Fig. \ref{fig:sfr_stellar_mass} also provides an insight into a possible evolutionary path for galaxies, progressing from the "main sequence" through the "green valley" and into the "non-star forming" region. If correct then this process must also be linked  to changes in morphology (Fig. \ref{fig:sfr_stellar_mass}). Thus this again, as in the chemical evolution model, implies a common evolutionary pathway and a process of changing morphology as a galaxy evolves. 

How the SSFR might be expected to change with time can be characterised by a model of the SF history. For example within the CIGALE SED fitting package the SF history is parameterised using the expression $SFR(t) \propto t\exp{-t/\tau}$, where $SFR(t)$ is the star formation rate at time $t$ and the scale factor $\tau$ governs the rate at which SF declines and the point at which it peaks\footnote{Derived values of $t$ and $\tau$ from CIGALE are discussed in more detail in Nersesian et al. (2018).}. Assuming this form for the SF history and that we again have closed box evolution (in this case that $M_{Baryon}=M_{Star}+M_{gas}$, which ignores the small contribution of metals in the dust and gas), this model predicts constant SSFR for galaxies that have the same value of $\tau$ and the ratio $t/\tau$ and does not depend on stellar mass. The actual relation is\footnote{The SFR is given by $SFR(t) \propto t\exp{-t/\tau}$ as stated above and $M_{Star}$ is the integral of the SFR from time=0 until time=$t$.}:
\begin{equation}
 \log{\frac{SFR}{M_{Stars}}}=-\log{\tau}+\log{\frac{1.0}{\frac{\tau}{t}(\exp{t/\tau}-1.0)-1.0}}
\end{equation} 
The observed "main sequence" (Fig. \ref{fig:sfr_stellar_mass}) actually has a shallow slope and so is not quite a line of constant SSFR\footnote{In some ways this is reminiscent of the main sequence of stars, which is not quite a line followed by a perfect blackbody.}. Within the bounds of this very simple SF history model (there is no "physics" of star formation in it) the position of the galaxies in Fig. \ref{fig:sfr_stellar_mass} is purely a consequence of their SF history (defined by $t$ and $\tau$). We will explore this further once we have considered the other substantial baryonic component, the gas mass.

Fig. \ref{fig:sfr_stellar_mass} also indicates that there is little or no difference between the SSFR and stellar masses of cluster and field galaxies. To quantify this further in Table \ref{table:specific_sfr} we give the mean values of SSFR$^{-1}$ (the time to form the current mass of stars at the current SFR). SSFR$^{-1}$ is essentially the same for the cluster and the field with the exception of the early type galaxies. On average Virgo cluster early type galaxies (E/S0) have lower values of SSFR than those in the field, but we need to be a little cautious with this interpretation. On average the Virgo early types are "earlier" than those in the field, while for the other morphological types there is no discernible difference in type between those in the cluster and field (Table \ref{table:morph}).

\begin{table}
\begin{center}
\begin{tabular}{c|c|c}
  Type & $<T_{Field}>$  &  $<T_{Virgo}>$  \\ \hline
E/S0 & -1.9$\pm0.2$ & -2.6$\pm0.2$  \\
Sa/Sb & 1.6$\pm0.1$ & 1.7$\pm0.2$  \\
Sc/Sd & 4.6$\pm0.1$ & 4.9$\pm0.2$ \\
Sm/Irr & 8.0$\pm0.2$ &  7.9$\pm0.3$ \\
\end{tabular}
\caption{Mean morphological types.}
\label{table:morph}
\end{center}
\end{table} 

\subsubsection{The star formation efficiency}
\begin{figure}
\centering
\includegraphics[scale=0.33]{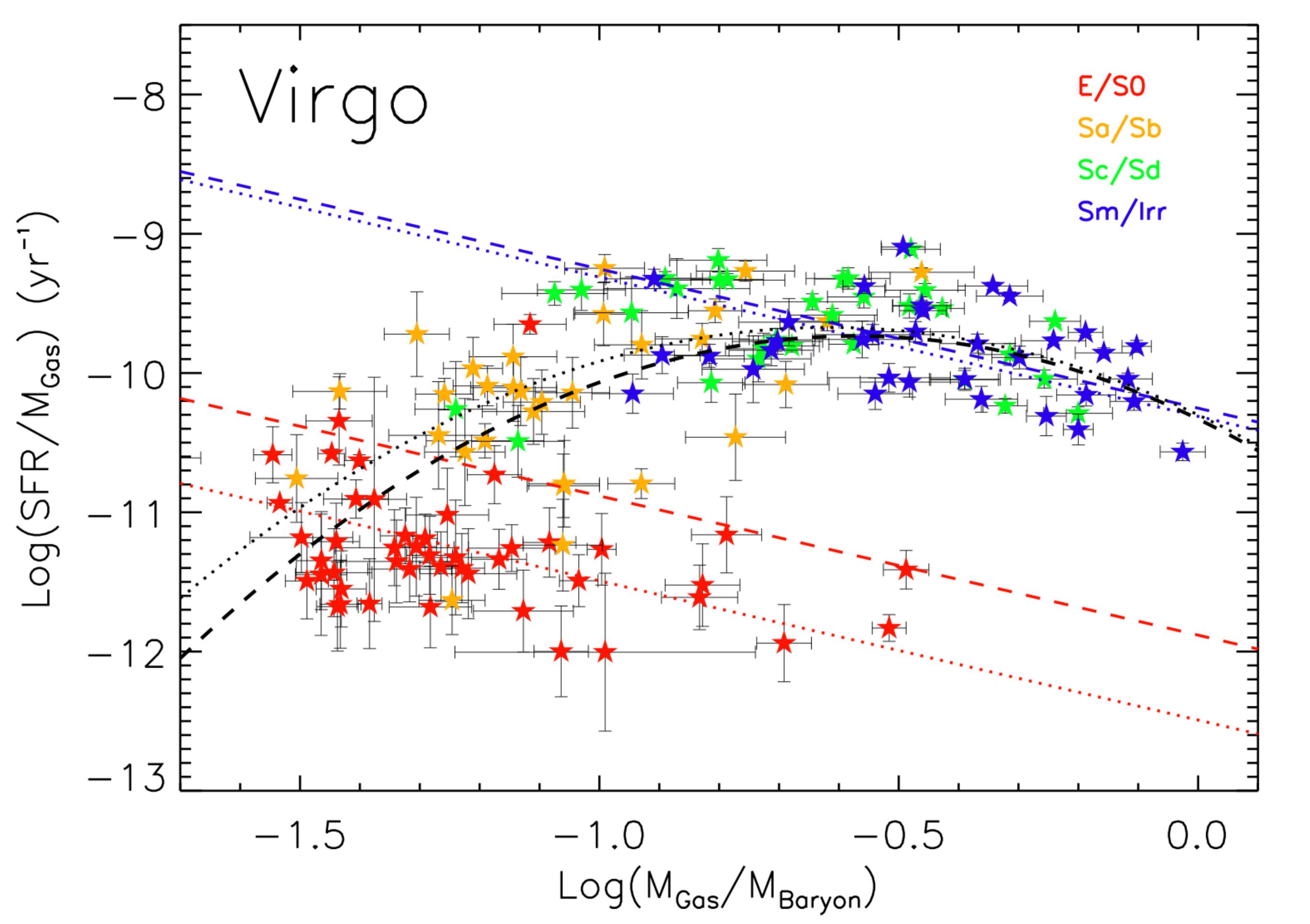}
\includegraphics[scale=0.33]{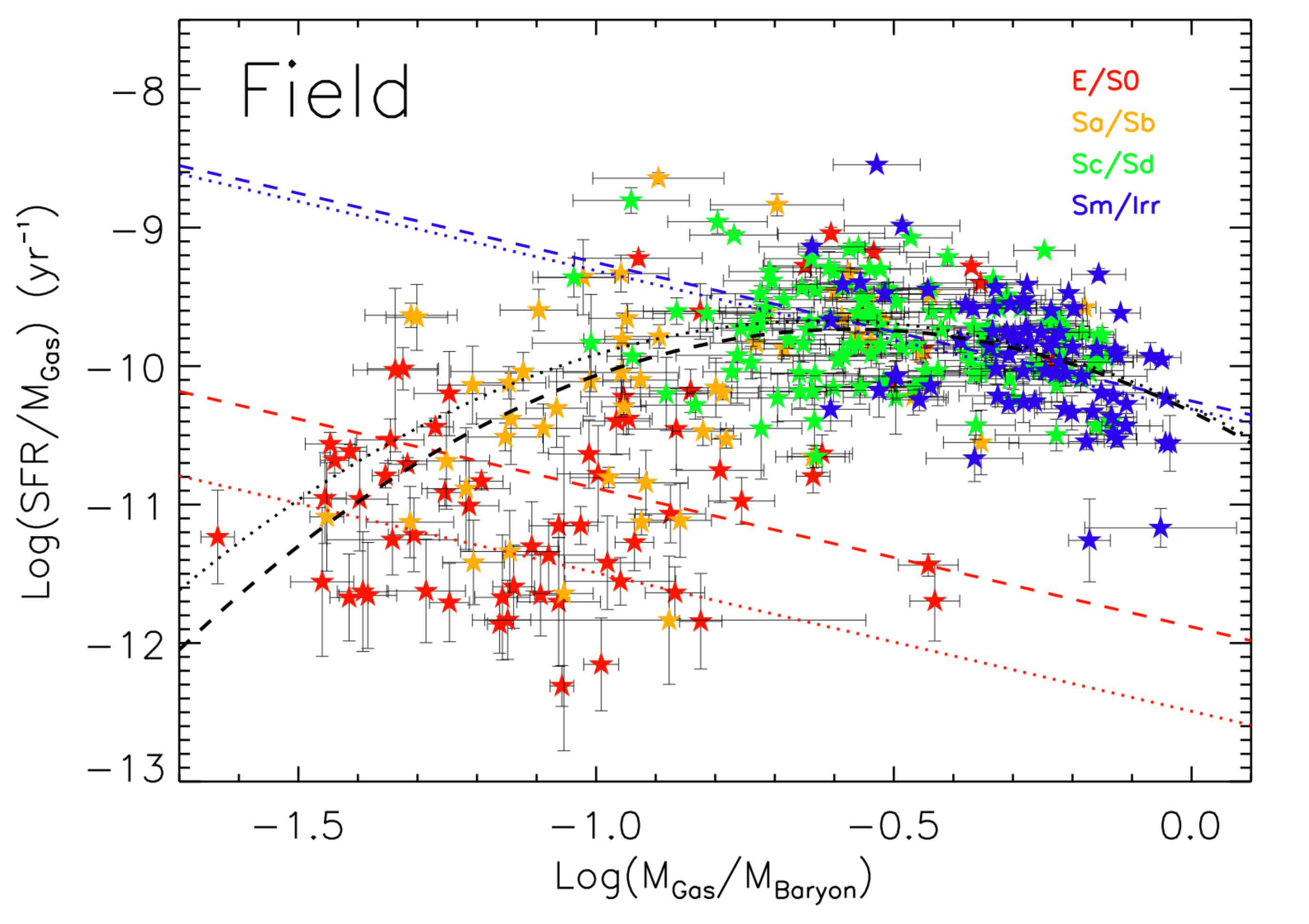}
\vspace{-0.2cm}
\caption{The star formation efficiency (SFR$/M_{Gas}$) versus the gas fraction ($M_{Gas}/M_{Baryon}$) for galaxies of different morphological types and both within the Virgo cluster (top) and the field (bottom). The blue and red lines are linear fits of slope -1 to the late (Sc/Sd and Sm/Irr) and early (E/S0) type galaxies for Virgo (dotted) and Field (dashed) respectively. The black curved lines are second order polynomial fits to the late types, in this case later than Sa, again Virgo (dotted) and Field (dashed).}
\label{fig:sf_eff}
\end{figure}

A possibly more interesting and physically relevant quantity than the SSFR, is the SFR normalised by the gas rather than stellar mass (the SFR per unit mass of gas, the SF efficiency or SFE). At first sight this is a measure of how efficiently stars are being produced from the gas at the current time and so may relate more directly to the physics of the star formation process. Here it is of interest because potentially the environment could affect how efficiently stars form. For example by providing additional gravitational interactions that enhance gas collapse into stars, provide additional gas for infall or strip gas such that there is a different relationship between the current SFE and the availability of gas.

In Fig. \ref{fig:sf_eff} we plot the SFE against the gas fraction. Clearly the efficiency of star formation varies with the availability of gas and also may take some time to "ramp up" once SF has begun and "ramp down" once star formation is in decline.  Looking at the data in Fig. 5 there is certainly the possibility of a peak in the data, which we have tried to quantify by fitting a second order polynomial to the data for types later than Sa (there is quite a lot of scatter in the data for galaxies earlier than this, Fig. \ref{fig:sf_eff}). The fitted line (black lines on Fig. \ref{fig:sf_eff}) are almost identical for field and cluster galaxies again indicating common internal physics governing the SF process. Interestingly the SFE seems to peak at a gas fraction of $f\approx0.25$ for field and cluster galaxies i.e. when about 75\% of the gas has been consumed. Note that according to the chemical evolution model the maximum dust fraction occurs for $f\approx0.37$  i.e. when $\Delta_{max,f}=\eta pf \ln{(1/f)}$ has its maximum value. 

That our prediction for the SFE is not constant is a natural consequence of the analytic expression previously used to define the SF history of a galaxy - it does not explicitly relate to physical processes in the interstellar medium.  It is straight forward to show that galaxies should lie on a straight line of slope -1 on Fig. \ref{fig:sf_eff} if they have the same "combination" of $t$ and $\tau$; for example if they started to form stars at the same time and with the same scale factor\footnote{The actual time derived from the intercept of the line is $\frac{\tau^{2}}{t\exp{-t/\tau}}$.}. In this case the SFE depends on when we observe the galaxy in relation to when it started forming stars. It is clearly possible that the SF galaxies (Sc/Sd and Sm/Irr) lie on such a line (blue lines, Fig. \ref{fig:sf_eff}) and so have common SF histories. Below we will derive a value for $t$ and $\tau$ for these star forming galaxies.

Combining the SF history with the closed box model leads to two interpretations of the SFE. Firstly, it is a measure of how efficiently gas is currently being converted into stars and so is directly related to physics in the ISM. Secondly, it relates to the timescale of star formation ($\tau$) and at what time ($t$) after star formation has begun  that we observe the galaxy. The second interpretation  probably also relates to physics in the ISM, but in a less direct way via the processes that drive the SF history.

The linear fits for the later types are almost identical for cluster and field galaxies (blue dotted and dashed lines on Fig. \ref{fig:sf_eff}) suggesting a common age and SF history profile irrespective of whether they belong to the field or the cluster. For the early types the linear fits are offset from each other in cluster and field indicating an older characteristic age for the cluster galaxies, but there is obviously much more scatter in the positions of the early type field galaxies in Fig. \ref{fig:sf_eff}, so much so that a linear fit of slope -1 is really not justified at all. As before we should also note the differences in mean morphological type between cluster and field early type galaxies (Table \ref{table:morph}).

\begin{table}
\begin{center}
\begin{tabular}{c|c|c}
  Type & $<-\log{(<SFE>)}_{Field}$ & $-\log{(<SFE>)}_{Virgo}$   \\
   &  ($yr$) & ($yr$) \\ \hline
E/S0 & $10.8\pm0.1$ & $11.3\pm0.1$ \\
Sa/Sb & $10.2\pm0.1$ & $10.2\pm0.1$ \\
Sc/Sd & $9.8\pm0.1$ & $9.7\pm0.1$ \\
Sm/Irr & $9.9\pm0.1$ & $9.9\pm0.1$ \\
\end{tabular}
\caption{The mean time to consume the gas at the current SFR for different morphological types both within the Virgo cluster and in the field.}
\label{table:sfeff}
\end{center}
\end{table} 

In summary, the current SFE of a galaxy is far from constant, but varies with the gas fraction and hence systematically with morphology. We find little or no difference in the relationship between SFE and gas fraction for galaxies in the field and in the cluster, indicating very little interference in the SF process by the environment. In a similar way to the the SSFR of the previous section, the mean times to consume the gas at the current star formation rate are consistent between cluster and field with the exception of the early type galaxies (E/S0), Table \ref{table:sfeff}. 

\subsubsection{The stars-to-dust mass ratio}
\begin{figure}
\centering
\includegraphics[scale=0.26]{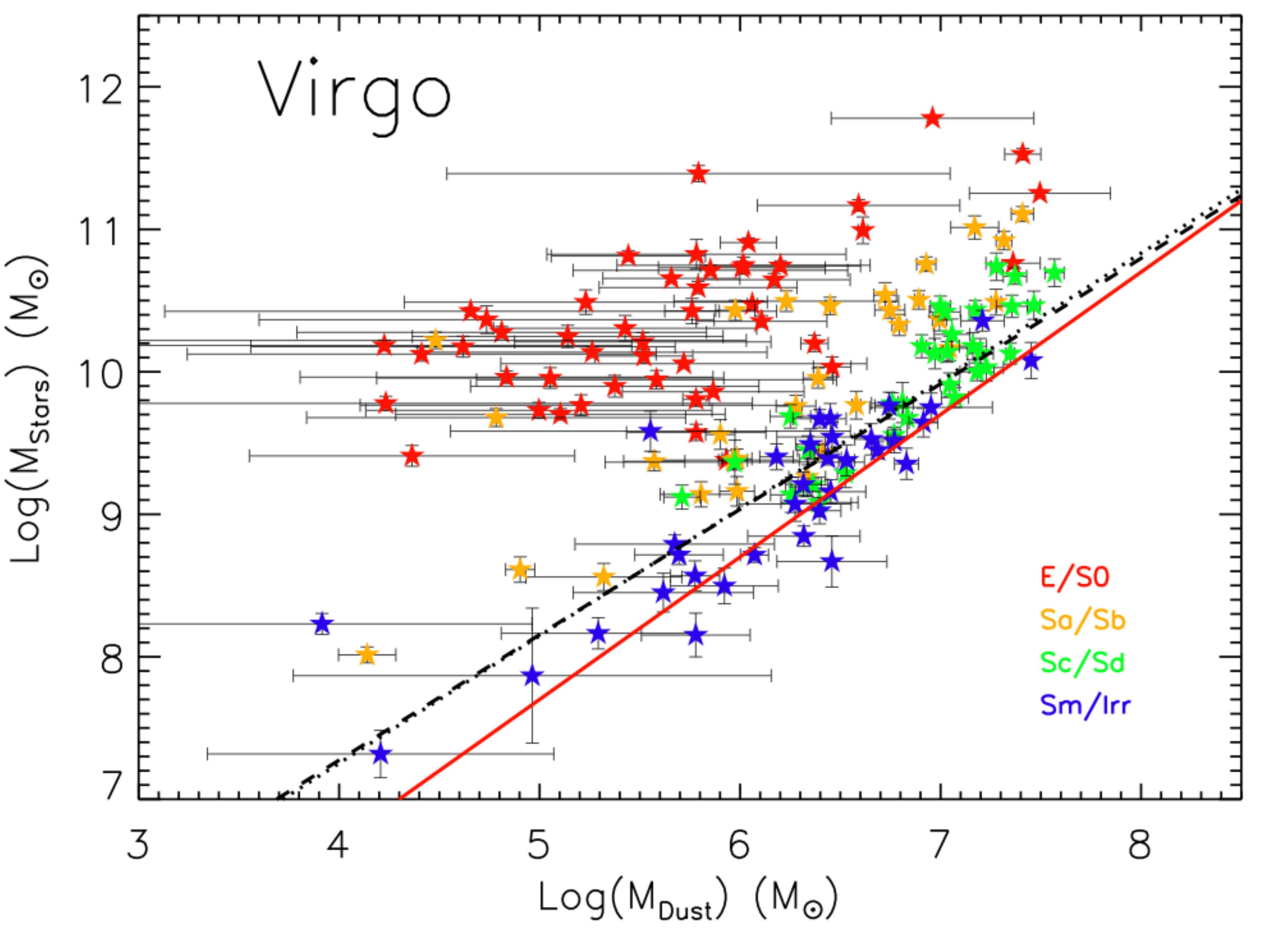}
\includegraphics[scale=0.26]{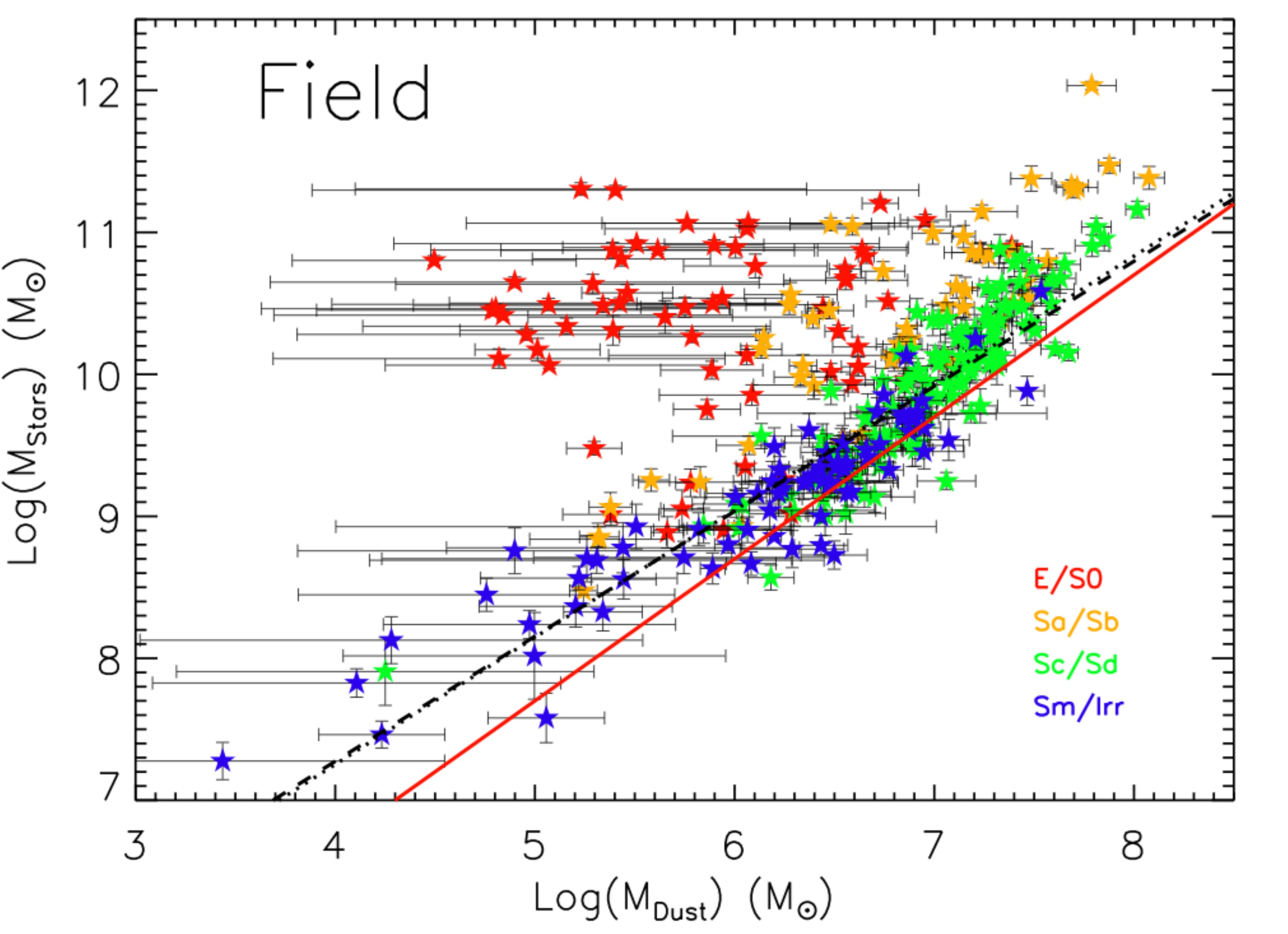}
\vspace{-0.2cm}
\caption{The relationship between stellar and dust mass (M$_{\odot}$) as both a function of morphological type and environment - Virgo (top) and Field (bottom). The dotted line is a fit to the Virgo and the dashed line a fit to the field data for morphological types Sc/Sd and Sm/Irr (the two lines sit on top of each other and so are difficult to distinguish). The red line corresponds to our calculated minimum value of $\log{(M_{Star}/M_{Dust})}=2.7$.}
\label{fig:star_dust_mass}
\end{figure}

 In Fig. \ref{fig:star_dust_mass} we plot the stellar mass against the dust mass. Although it is not generally very informative to plot extensive properties like this (things big in one component tend to be big in another), here we think this plot is informative as it illustrates common properties over a large range of intrinsic masses. From Fig.  \ref{fig:star_dust_mass} we see that typically earlier type galaxies have higher values of $M_{Star}/M_{Dust}$, but that there is a lot of scatter, particularly for the earliest types (E/S0). The later types (Sc/Sd and Sm/Irr) lie on an approximately linear relation, which is almost identical for galaxies in the cluster and in the field (see also Cortese et al., 2012). If we fix the slope of the fitted line at unity then both cluster and field star forming galaxies lie on a line with an almost constant star-to-dust mass ratio of $10^{3}$ (actual values are $10^{3.07}$ for Virgo and $10^{3.08}$ for the field). For these later type galaxies (Sc/Sd and Sm/Irr) the internal physical process of SF and subsequent chemical evolution have led to an almost constant star-to-dust ratio over a range in stellar mass of $10^{4}$ and this appears to be independent of the environment.
 
To further quantify this in Table \ref{table:star_dust}  we tabulate the mean values of $M_{Star}/M_{Dust}$ for galaxies of different types in the cluster and field environments. For both cluster and field samples values of $M_{Star}/M_{Dust}$ decrease  towards later morphological types, but there is no discernible difference between $M_{Star}/M_{Dust}$ for the cluster and field samples. 

\begin{table}
\begin{center}
\begin{tabular}{c|c|c}
  Type & $\log{(<M_{Star}/M_{Dust}>)_{Field}}$ &  $\log{(<M_{Star}/M_{Dust}>)_{Virgo }}$  \\ \hline
E/S0 & $4.5\pm0.1$ & $4.7\pm0.1$    \\
Sa/Sb & $3.6\pm0.1$ & $3.7\pm0.1$    \\
Sc/Sd & $3.0\pm0.1$ & $3.1\pm0.1$    \\
Sm/Irr & $3.0\pm0.1$ & $2.9\pm0.1$    \\
\end{tabular}
\caption{Mean values of $M_{Star}/M_{Dust}$ for field and cluster galaxies of different morphological types.}
\label{table:star_dust}
\end{center}
\end{table} 

We can interpret the data shown in Fig. \ref{fig:star_dust_mass} by combining the chemical evolution model and star formation history discussed in the previous sections. Combining these two models leads to the prediction that the stars-to-dust mass ratio is given by: 

$\frac{M_{Stars}}{M_{Dust}}=\frac{1}{\eta p}\frac{x-1}{x\ln{x}}$ 
where $x=(1+(t/\tau))\exp{-t/\tau}$.
The expression $\frac{x-1}{x\ln{x}}$ tends to unity for small $t$ and so predicts a minimum value of $\frac{M_{Stars}}{M_{Dust}}=\frac{1}{\eta p}=500$ or $\log{(1/\eta p)}=2.7$ using the value of $\eta p\approx2.0\times10^{-3}$ derived above. This minimum value is illustrated by the red line on Fig.  \ref{fig:star_dust_mass} and within the errors on each data point is consistent with the observational data. Our value is consistent with previous estimates of this number i.e. $\log{(M_{Star}/M_{Dust})_{min}} \approx 2.5$ (Dunne et al., 2011, Edmunds and Eales, 1998).

The function $\frac{1}{\eta p}\frac{x-1}{x\ln{x}}$ varies quite slowly with increasing values of $t/\tau$ (black dotted line Fig. \ref{fig:sf_hist}). For $\log{(\frac{M_{Stars}}{M_{Dust}})} \approx 3.0$ (as observed) $t/\tau \approx 2.0$ - the SFR peaks approximately half way through the star forming life of a galaxy. This provides an explanation of why so many of the galaxies shown in Fig. \ref{fig:star_dust_mass} have values of $\log{(M_{Stars}/M_{Dust})}$ close to 3.0, see also Table \ref{table:star_dust}. 

Values of $\log{(\frac{M_{Stars}}{M_{Dust}})}$>4.0 for early types (Table \ref{table:star_dust}) are consistent with a peak SFR much closer to the time of formation i.e. values of $t/\tau \approx 7$ and above (Fig. \ref{fig:sf_hist}). However, there is also a difficult to interpret, within the bounds of our simple model, wide range in observed dust mass for a given stellar mass.

In summary the data relating stellar and dust mass for late type galaxies (Sc to Irr) are consistent with and explainable by the simple chemical evolution model combined with the star formation history. There are no measurable differences between galaxies residing inside or outside of the Virgo cluster. The situation is not so straightforward for the early types (particulary E/S0). Mean values of $M_{Star}/M_{Dust}$ are consistent between cluster and field, but there is considerable scatter in the data, which is difficult to model within the current framework. 

\subsubsection{The gas-to-dust mass ratio}
\begin{figure}
\centering
\includegraphics[scale=0.34]{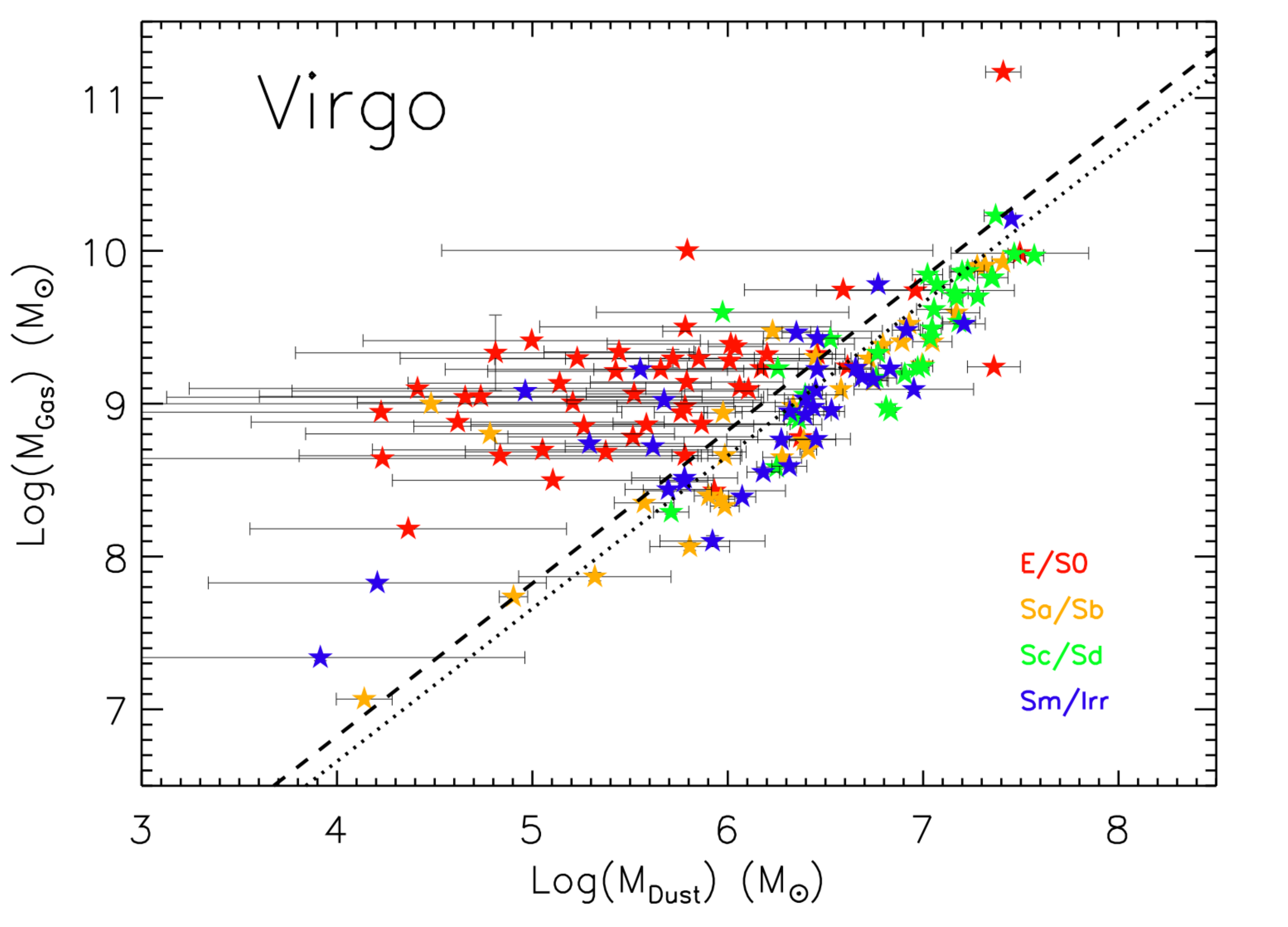}
\includegraphics[scale=0.34]{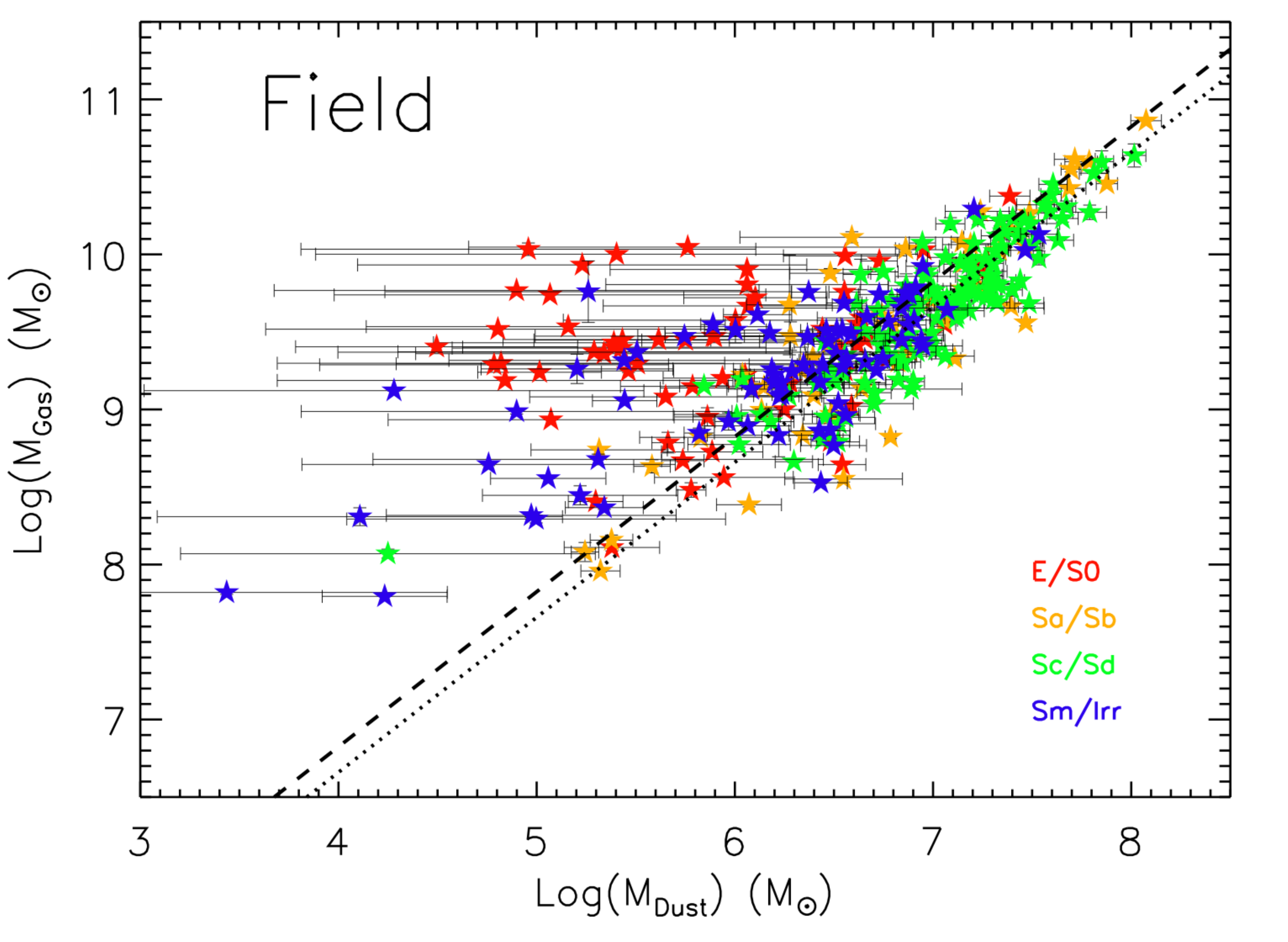}
\vspace{-0.2cm}
\caption{The relationship between total gas and dust mass (M$_{\odot}$) as both a function of morphological type and environment - Virgo (top) and Field (bottom). The dotted and dashed lines correspond to constant values of $\log{(M_{Gas}/M_{Dust})}=2.7\pm0.1$ and $2.8\pm0.1$ for Virgo and field star forming galaxies (Sc/Sd and Sm/Irr) respectively.}
\label{fig:gas_dust_mass}
\end{figure}

In Fig. \ref{fig:gas_dust_mass} we show the relationship between $M_{Dust}$ and $M_{Gas}$ for cluster (top) and field (bottom) galaxies. Again for the later types (Sc/Sd, Sm/Irr) there is a well-defined relation between $M_{Dust}$ and $M_{Gas}$ that can be simply interpreted as an approximately  constant gas-to-dust ratio. A fit to the data gives a value of $\log{(M_{Gas}/M_{Dust})}=2.7\pm0.1$ and $2.8\pm0.1$ for cluster and field late type galaxies respectively (dotted and dashed lines on Fig. \ref{fig:gas_dust_mass}). 

There is an indication in the data (Fig. \ref{fig:gas_dust_mass}) that the gas-to-dust ratios in both cluster and field increase from this constant value for $\log{M_{Dust}}<6$. These same galaxies have the lowest stellar masses in the sample (Fig.  \ref{fig:star_dust_mass}) and the highest SSFR (Fig. \ref{fig:gas_dust_mass}) - they are low mass, gas rich, star forming galaxies that probably have not yet enriched their ISM with metals to the same extent as more massive galaxies. These galaxies are present in both the field and the cluster. As for the stellar mass (Fig.  \ref{table:star_dust}) early type galaxies can have a wide range of dust mass for a given gas mass. Generally Fig. \ref{fig:gas_dust_mass} and Table \ref{table:gas_dust} indicate that there is no apparent difference between cluster and field galaxies with regard to their gas-to-dust ratios.

\begin{table}
\begin{center}
\begin{tabular}{c|c|c}
  Type & $\log{(<M_{Gas}/M_{Dust}>)_{Field}}$ &  $\log{(<M_{Gas}/M_{Dust}>)_{Virgo }}$  \\ \hline
E/S0 & $3.5\pm0.1$ & $3.5\pm0.1$    \\
Sa/Sb & $2.8\pm0.1$ & $2.7\pm0.1$    \\
Sc/Sd & $2.7\pm0.1$ & $2.6\pm0.1$    \\
Sm/Irr & $3.1\pm0.1$ & $2.8\pm0.1$    \\
\end{tabular}
\caption{Mean values of $M_{Gas}/M_{Dust}$ for field and cluster galaxies of different morphological types.}
\label{table:gas_dust}
\end{center}
\end{table} 

We can again interpret the data shown in Fig. \ref{fig:gas_dust_mass} using the chemical evolution model and star formation history discussed in the previous sections. In this case the gas-to-dust mass ratio is given by: 
$\frac{M_{Gas}}{M_{Dust}}=\frac{-1}{\eta p \ln{x}}$.

\begin{figure}
\centering
\includegraphics[scale=0.24]{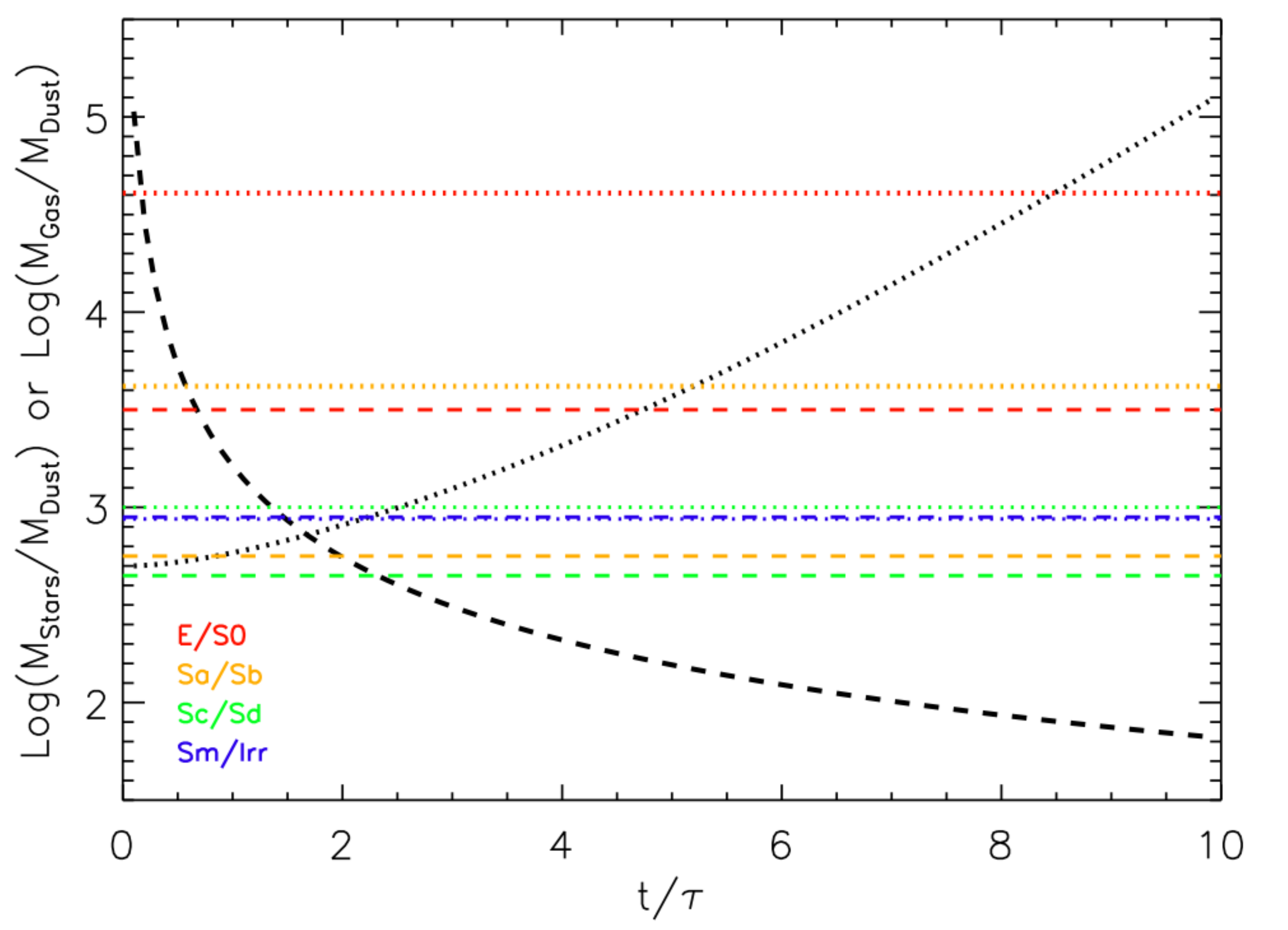}
\caption{Model predictions for  $M_{Stars}/M_{Dust}$ (dotted, black) and $M_{Gas}/M_{Dust}$ (dashed black) ratios as a function of the SFR parameter $t/\tau$, using a value of $\eta p=2.0 \times 10^{-3}$. The mean observed values of $M_{Stars}/M_{Dust}$ (dotted) and $M_{Gas}/M_{Dust}$ (dashed) for galaxies of different morphological types are indicated by the coloured lines.}
\label{fig:sf_hist}
\end{figure}

The expression $\frac{-1}{\eta p \ln{x}}$ tends to zero for large $t/\tau$ but is quite flat over the interval $1.0<t/\tau<8.0$ (black dashed line Fig. \ref{fig:sf_hist}), which again using a value of $\eta p=2.0\times 10^{-3}$ embraces the range of values for $\frac{M_{Gas}}{M_{Dust}}$ given in Table \ref{table:gas_dust} and that derived from the linear relation shown in Fig. \ref{fig:gas_dust_mass}. 

The mean gas-to-dust mass ratio of the field and cluster samples is the same (both have $\log(<M_{Gas}/M_{Dust}>)=2.95\pm0.04$). If cluster galaxies are affected by gas stripping processes, for example ram pressure stripping, then the dust seems to have been stripped as well, as might be expected because of their close collisional coupling.

Given the above discussions of the $M_{Stars}/M_{Dust}$ and $M_{Gas}/M_{Dust}$ ratios we now consider whether these two ratios lead to consistent values of the star formation history parameter $t/\tau$. In Fig. \ref{fig:sf_hist} we show the model predictions for these values (black lines). As the values of $M_{Stars}/M_{Dust}$ and $M_{Gas}/M_{Dust}$ do not vary significantly between cluster and field (Tables  \ref{table:star_dust} and  \ref{table:gas_dust}) we consider here the combined cluster and field samples. 

The coloured lines on Fig. \ref{fig:sf_hist} show the observed mean values of $M_{Stars}/M_{Dust}$ and $M_{Gas}/M_{Dust}$ (dotted and dashed respectively) for the different morphological types. For consistency with the model we should expect that a dotted horizontal line should cross the model dotted line at the same value of $t/\tau$ as the dashed horizontal line crosses the dashed model line i.e. both $M_{Stars}/M_{Dust}$ and $M_{Gas}/M_{Dust}$ provide a consistent value for $t/\tau$. 

For example the blue dashed and dotted lines (Sm/Irr) lie almost on top of each other ($M_{Stars}/M_{Dust} \approx M_{Gas}/M_{Dust}$) and they cross the model (black) dashed and dotted lines at about the same value of $t/\tau$ - interesting this is just where the model predicts that $M_{Stars}/M_{Dust} = M_{Gas}/M_{Dust}$. In this case the observations are perfectly consistent with the model. The same is approximately true for Sc/Sd galaxies. The green dashed line crosses the black dashed line at just about the same value of $t/\tau$  as the green dotted line crosses the black dotted line - consistent with the closed box model and the SF history model we have used. 

For these late type galaxies the consistent value is $t/\tau \approx 2$. Using eq. 1 and a value of $\log{SFR/M_{Stars}} \approx -10.0$ (Table \ref{table:specific_sfr}) leads to values of $\tau=4.6 \times 10^{9}$ and $t=9.2 \times 10^{9}$ years. With these values SF would start at a redshift of $z\approx1.4$ and peak at a redshift of $z\approx0.43$, given the cosmological model (Ade et al. 2016). For these late type galaxies SFRs, stellar, gas and dust masses are consistent with the simple model of how the SFR varies with time and closed box chemical evolution. Importantly for this paper the model and timescales are the same irrespective of whether the galaxies reside in the cluster or the field - within the bounds of this simple model there is no environmental effect on these late type galaxies. 

However, looking a little more closely and considering the $M_{Stars}/M_{Gas}$ ratio something different between the two environments does become apparent. The global stars-to-gas mass ratio does show a marked difference between the cluster and field - $\log(<M_{Stars}/M_{Gas}>)_{Virgo}=0.77\pm0.05$, $\log(<M_{Stars}/M_{Gas}>)_{Field}=0.41\pm0.03$ and as Table \ref{table:star_gas} shows this is not just due to differences in the morphological mix. The reason for this is that values of $M_{Stars}/M_{Dust}$ are slightly higher in the cluster compared to the field (Table \ref{table:star_dust}) and values of $M_{Gas}/M_{Dust}$ are slightly lower (Table \ref{table:gas_dust}), though neither on its own is significant within the errors. When combined this leads to a measurable  difference  in the values of $M_{Stars}/M_{Gas}$ between cluster and field. This is the well known gas depletion of Virgo galaxies compared to the field (Giovanelli and Haynes, 1985). It is interesting that in all other respects measured here there is no difference between the properties of the Virgo and field late type galaxies. This is something that Kennicutt (1983) has previously commented on and perhaps, as we said earlier, the Virgo galaxies are just a little "more evolved" rather than being stripped of their gas.

\begin{table}
\begin{center}
\begin{tabular}{c|c|c}
  Type & $\log{(<M_{Stars}/M_{Gas}>)_{Field}}$ &  $\log{(<M_{Stars}/M_{Gas}>)_{Virgo }}$  \\ \hline
E/S0 & $1.0\pm0.1$ & $1.2\pm0.1$    \\
Sa/Sb & $1.0\pm0.1$ & $1.0\pm0.1$    \\
Sc/Sd & $0.3\pm0.1$ & $0.5\pm0.1$    \\
Sm/Irr & $-0.1\pm0.1$ & $0.2\pm0.1$    \\
\end{tabular}
\caption{Mean values of $M_{Stars}/M_{Gas}$ for field and cluster galaxies of different morphological types.}
\label{table:star_gas}
\end{center}
\end{table} 

The correspondence between dashed and dotted lines (Fig. \ref{fig:sf_hist}) becomes progressively worse for the earlier types. Using $M_{Stars}/M_{Dust}$ (dotted line) the Sa/Sb types predict $t/\tau \approx 5$, while $M_{Gas}/M_{Dust}$ (dashed line) predicts $t/\tau \approx 2$. The situation for the early types is even worse at values $t/\tau \approx 8$ and 1 respectively. These early types clearly do not fit the model very well and so cannot be explained by the simple chemical evolution and SF history model. An obvious suggestion is that there is no longer closed box evolution and so we might infer that earlier types are increasingly (with T type) affected by either gas in-fall or mergers, a possible clue as to the origin of morphology.

The interesting conclusion is that whether galaxies fit the simple model or not, is not dependent on the environment, but depends on morphology irrespective of environment.

\subsubsection{Star formation and dust temperature}
There has been an ongoing debate about the importance of dust heating via the general interstellar radiation field compared to that produced by dust grains in close proximately to hot young stars (Viaene et al. 2016 and references therein). Whatever the final conclusion, there is clearly a close relationship between far infrared emission and star formation - with various measures of the far infrared luminosity being used as SFR measures (Davies et al. 2016b and references therein). 

\begin{figure}
\centering
\includegraphics[scale=0.25]{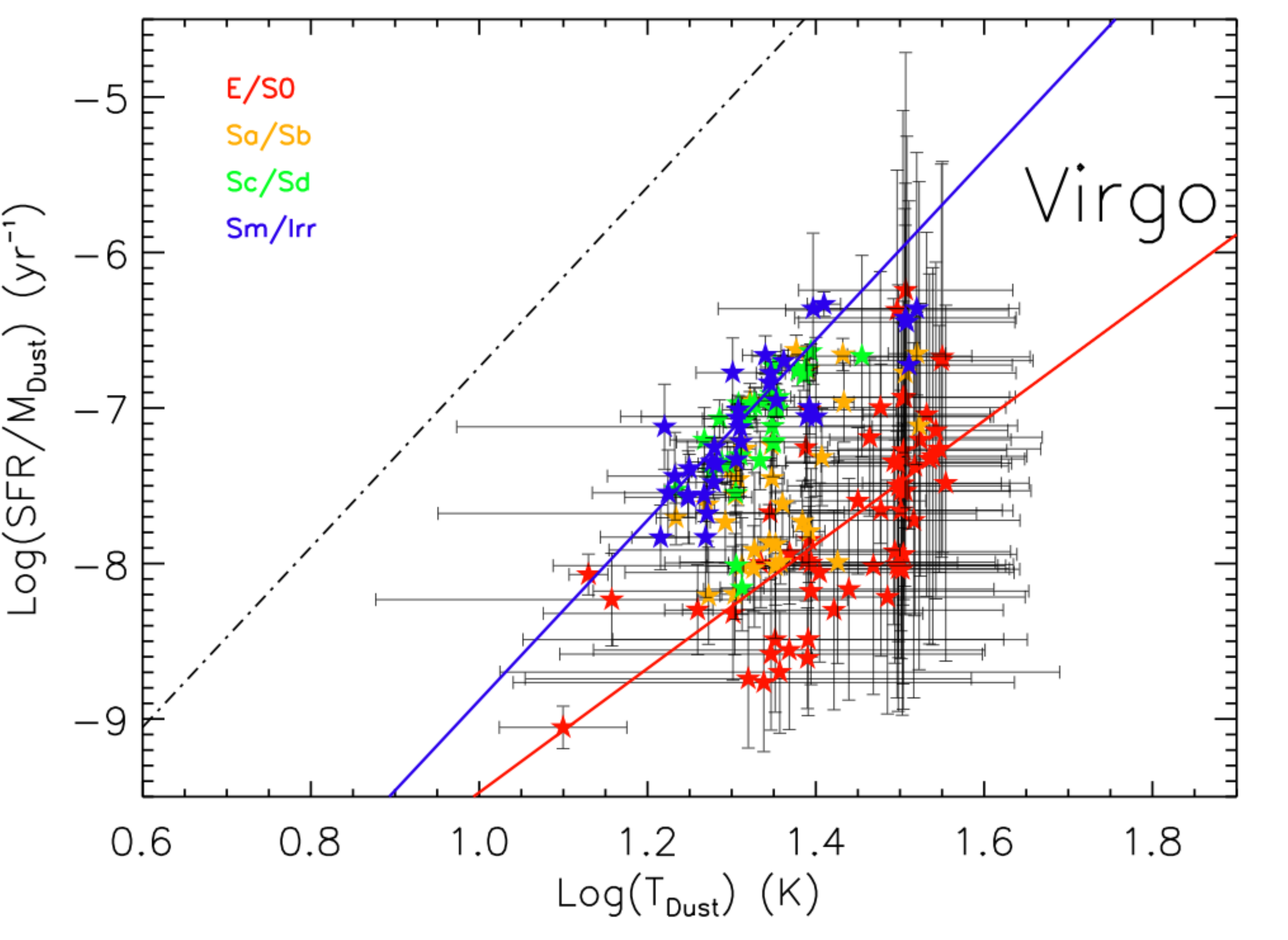}
\includegraphics[scale=0.25]{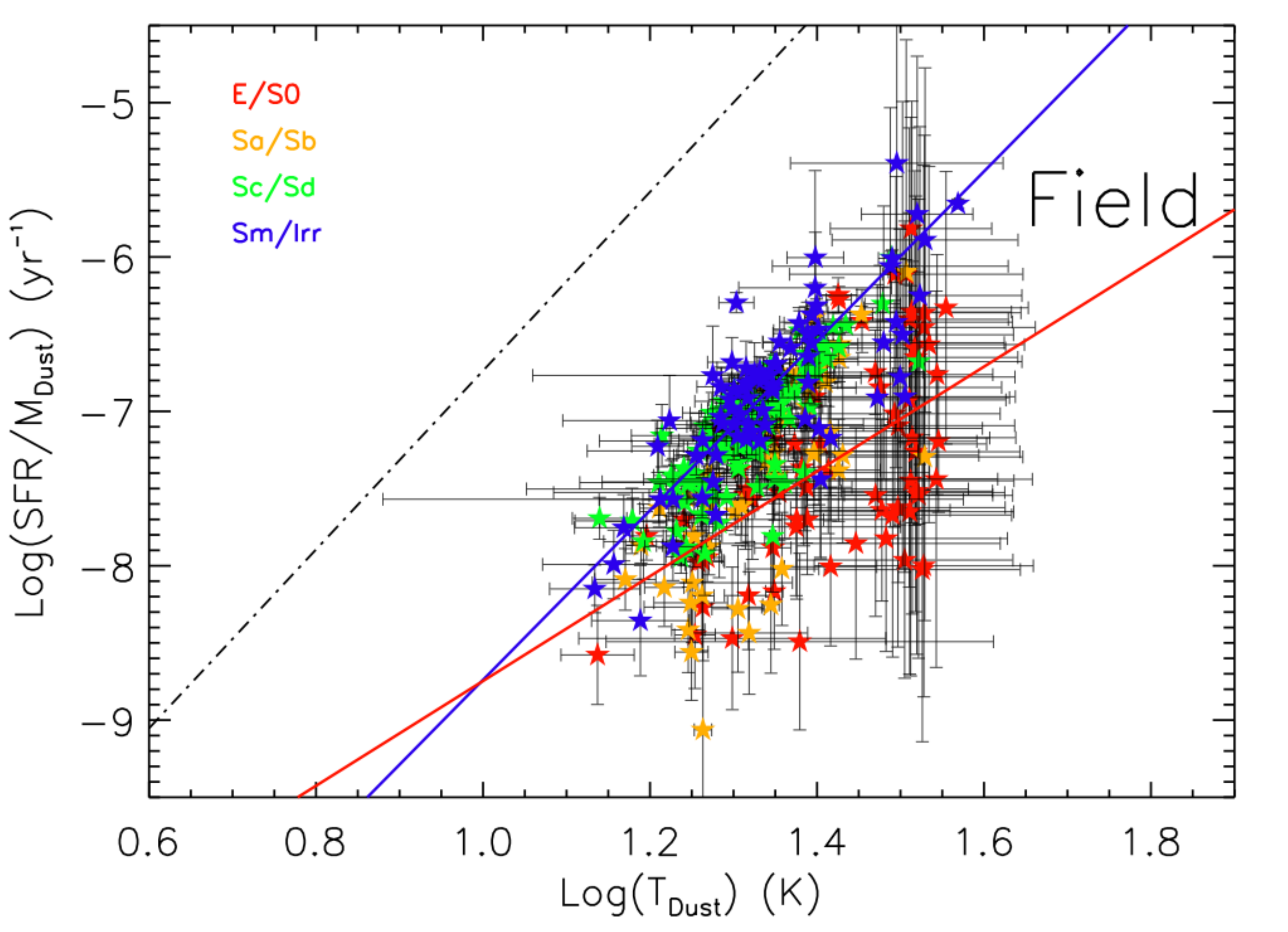}
\vspace{0.0cm}
\caption{The relationship between star formation rate per unit dust mass ($yr^{-1}$) and the dust temperature (K) as both a function of morphological type and environment - Virgo (top) and Field (bottom). The dot-dashed line illustrates the slope (not intercept) of the expected relationship if dust heating is primarily driven by star formation. The blue line is a linear fit to the late types (Sc/Sd and Sm/Irr) and the red line to the early types (E/S0).}
\label{fig:sfr_temp_ratio}
\end{figure}

In Fig. \ref{fig:sfr_temp_ratio} we show the relationship between the measured dust temperature and the dust specific SFR.  We can use this plot to investigate the fraction of heating contributed by the general interstellar radiation field and that due to star formation, there are two limiting cases. Firstly, if the dust is heated by the general interstellar radiation field then we should find no relationship between the dust temperature and the dust specific SFR. Secondly, if  the dust is heated entirely by young stars in star forming regions then the SFR per unit mass of dust is proportional to the total far infrared luminosity and $L_{FIR} \propto T_{Dust}^{4+\beta}$, where $\beta$ is the dust emissivity (Clemens et al., 2013). Thus we would expect the gradient of a line fitted to the data shown in Fig. \ref{fig:sfr_temp_ratio} to be zero for the first case and 5.79 (THEMIS model, dot-dashed line Fig. \ref{fig:sfr_temp_ratio}) for the second. $\beta$ for the THEMIS model has been obtained by fitting a modified blackbody curve to the predicted SED (Nersesian et al., 2018).

For the more actively star forming galaxies (Sc/Sd and Sm/Irr) we measure gradients of $5.8\pm0.7$ and $5.5\pm0.2$ for Virgo and field respectively (Fig. \ref{fig:sfr_temp_ratio}). The gradients are steep and consistent with the major source of dust heating being young stars. The situation is quite different for the early type galaxies (E/S0) where the relationship between dust specific SFR and temperature has more scatter and is far less steep - it is therefore more likely to be due to heating from sources other than SF (red line Fig. \ref{fig:sfr_temp_ratio}). Thus the issue of the source of dust heating is morphological type dependent and we can conclude that the small amount of dust in early type galaxies is not being heated by the small levels of SF, but by something else. X-ray and electron heating could play an important role (Goudfrooij and de Jong, 1995). However, given the reasonably good SED matches CIGALE obtained for these galaxies, which does not include X-ray and electron heating, predominately heating by the massive population of old stars appears to be most likely (De Vis et al. 2017a). Heating by the old stellar population appears to be increasingly important as one goes from later to earlier types, consistent with the decline if star formation and the increase in the mass of older stars.

We find that on average the dust is hotter in the early (E/S0) type galaxies (Table \ref{table:temp}), something previously found by Smith et al. (2012). At first sight this is counter intuitive because these are just the galaxies that do not appear to have dust heated by hot young stars. However, the dust temperatures we use here are derived from the mean intensity of the starlight as derived as part of the CIGALE SED fitting process\footnote{These are not the same as the more often used dust temperatures obtained from fitting a modified blackbody to the far infrared SED} and so are consistent across the SED with heating by stars only.  A full discussion and comparison of dust temperatures derived in different ways will be given in Nersesian et al., (2018). 

Differences in dust temperature between galaxies appear to be morphologically type dependent and not environmentally dependent (Table \ref{table:temp}).  

\begin{table}
\begin{center}
\begin{tabular}{c|c|c}
  & $<T_{Dust}>_{Field}$ & $<T_{Dust}>_{Virgo}$   \\ \hline
E/S0 & $26.8\pm0.7$ & $28.0\pm0.8$ \\
Sa/Sb & $22.0\pm0.5$ & $23.2\pm0.7$ \\
Sc/Sd & $21.1\pm0.3$ & $21.6\pm0.4$ \\
Sm/Irr & $22.8\pm0.6$ & $21.8\pm0.8$ \\
\end{tabular}
\caption{The median dust temperatures ($K$) of galaxies of different morphological types both within the Virgo cluster and in the field.}
\label{table:temp}
\end{center}
\end{table} 

\subsection{A comparison between galaxies in different environments}
In this section we will describe and compare the properties of the galaxies discussed in the previous sections, but now in relation to their local environment. We will define the environment of a particular galaxy by measuring the density of SDSS galaxies around it. Muldrew et al. (2012) detail different methods to estimate the local density. These methods fall into two groups, though the parameters previously used within these two groups can vary quite widely. 

The first method is to count nearest neighbours and define a density using $\sigma_{n}=\frac{n}{\pi r^{2}_{n}}$, where $n$ is the number of neighbours and $r_{n}$ the radius to the $n$th nearest neighbour. This is often used to derive a surface density, but can be used to obtain a volume density if good distance information is available. Various values of $n$ have been used in the past (see Muldrew et al. 2012, for a review). A variation is to add a velocity criteria on the selection of nearest neighbours such that the velocity difference $(\Delta v)$ between a galaxy and its neighbour is less than some value. 

The second method involves using a fixed physical sized aperture and count galaxies that reside within and then define a density contrast as $\delta=\frac{N_{g}-\bar{N}_{g}}{\bar{N}_{g}}$, where $\bar{N}_{g}$ is a normalising number density, such as, for example, the number expected for a random distribution. Again this can be carried out using volume or surface density and possibly the use of an additional velocity difference criteria. 

Muldrew et al. (2012) look at and compare twenty different methods of defining the environment and conclude that the nearest neighbour method is best used if the intention is to investigate the local environment internal to an individual halo, but that the fixed aperture method is best suited to investigations of the 'larger-scale environment'. As our intention here is to investigate the effect of the larger scale structure, but still local to each galaxy (see below) and not the specific structure of individual halos we will use the second method and use a fixed aperture and define a density contrast. 

We will employ a similar method to Gallazzi et al. (2009) in which they count the number of galaxies within a fixed circular aperture of radius $1h^{-1}\approx$1.5 Mpc (for $H_{0}=67.8$, Ade et al. 2016), but only count galaxies that also have $\Delta v < \pm 500$ km s$^{-1}$ to try and eliminate line-of-sight interlopers. This distance scale is consistent with the conclusion of Park et al. (2008) that companions within Mpc scales have the most important effect on morphology, star formation and luminosity (see also Scudder et al. 2012, Davies et al. 2015, 2016a and an application to numerial simulations in Wang et al. 2018). However, we consider that the value of $\Delta v$ Gallazzi et al. (2009) use is too large. For example at the median velocity of the DustPedia sample (1386 km s$^{-1}$) $\pm500$ km s$^{-1}$ corresponds to $\approx$15 Mpc if due to the Hubble expansion, and the detection volume in this case, is a long thin cylinder. 

To make a more reasonable estimate of what $\Delta v$ should be, we have considered the Local Group of galaxies. The median $g$ band apparent magnitude of the SDSS data is 16.6, which at the median distance of 20.0 Mpc gives an absolute magnitude of -14.9. Using the data in McConnachie (2012) (and assuming $(g-V) \approx 1.0$) we can estimate the number of Local Group galaxies that would be detected in our SDSS data if centred on the Milky Way. There are just six galaxies sufficiently bright and within a 1.5 Mpc aperture radius (M31, M33, LMC, SMC, M32 and NGC205). The standard deviation of the line-of-sight velocities of these six galaxies is 238 km s$^{-1}$ and so this is the value we will use as our velocity selection i.e. $\Delta v \le 238$ km s$^{-1}$. This velocity is typical of the relative velocities found for galaxies in groups and roughly corresponds to the typical rotational velocity of a galaxy like the Milky Way. Galaxies have the most influence on each other when their approach velocity is of order their rotation velocity and so the most influential galaxies will have $\Delta v$ of order this value. This value of $\Delta v$ also now gives a more comparable value for the line-of-sight detection length (if in the Hubble expansion at the median distance) of $\sim7.0$ Mpc. 

As a measure of the environment we use the density contrast as defined above with the arbitrary constant $(\bar{N}_{g}$) set to a value of 6.83 (the number of Mpc$^{2}$ in the detection aperture). Normalised in this way a value of $\delta=0.0$ corresponds to a surface density of one galaxy per Mpc$^{2}$. When $\delta=-0.85$ there is just one galaxy in the detection volume i.e. a galaxy that should not have been affected by its environment because it has no neighbours. With a fixed physical length aperture the angular size varies with galaxy distance from us. On Fig. 1 we show the angular size of a 1.5 Mpc radius aperture at the median distance of the DustPedia sample used here (20.0 Mpc). 

The SDSS data we use covers an area of about 10$^{o}$ larger on a side than that used by us to select the DustPedia sample (Fig. 1) and our velocity selection for SDSS galaxies extends to 3500, rather than 3000 km s$^{-1}$. We have done this so that we avoid edge effects. For nearby galaxies this border becomes an issue (too small) when the radial size of an individual aperture is of order 10$^{o}$. However, there are just 25 galaxies that have apertures with radial sizes larger than 10$^{o}$ and only 10 of these extend outside the SDSS area to varying degrees and we have left them in the sample. 

\begin{figure}
\centering
\includegraphics[scale=0.26]{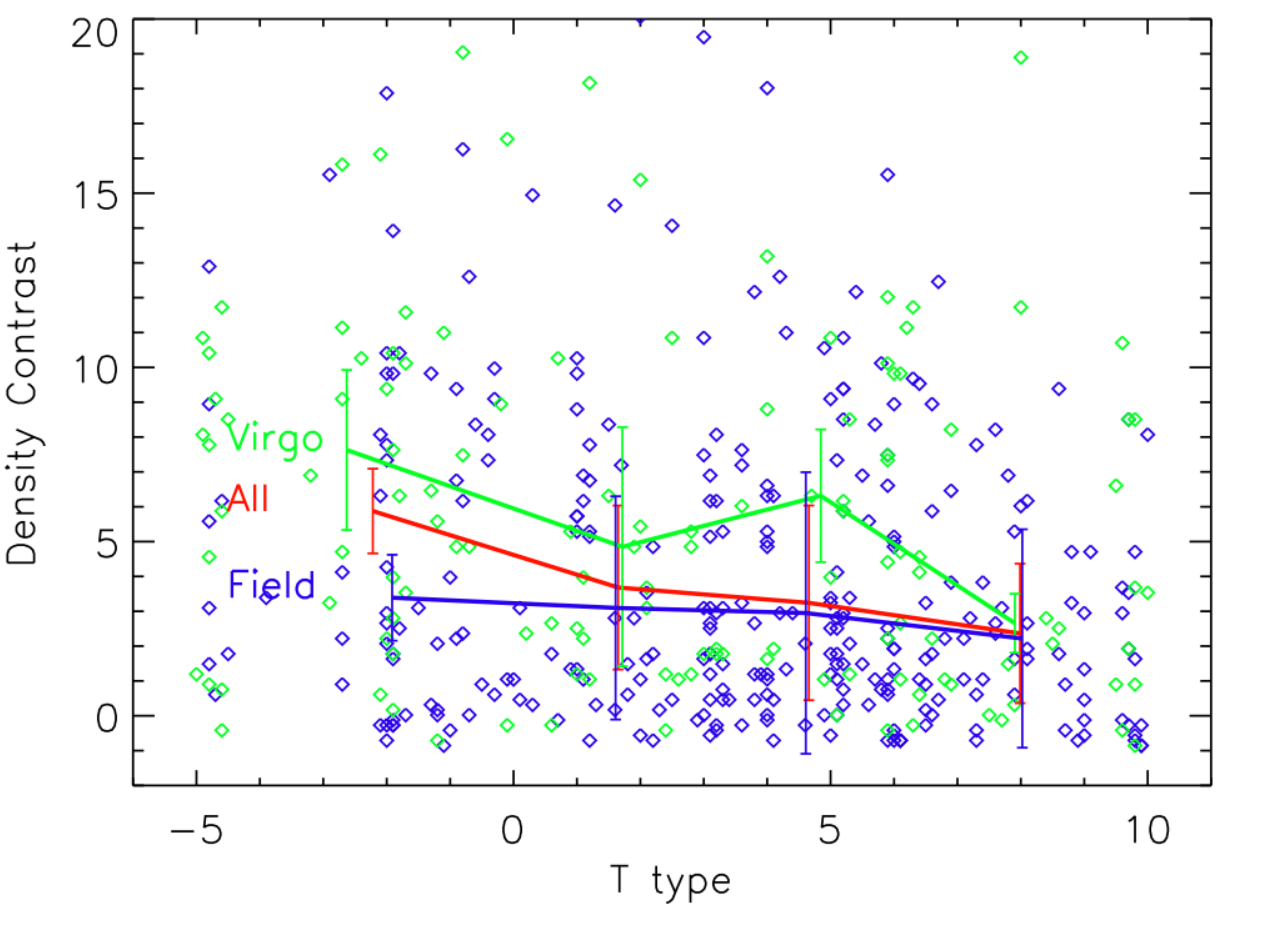}
\vspace{-0.3cm}
\caption{The distribution of the density contrast parameter ($\delta$) as a function of galaxy type. Blue symbols are for galaxies outside of the Virgo cluster and green for those within the Virgo virial radius. The red line connects the median values of $\delta$ for the morphological types  defined in Table 1.  The green line connects median values for galaxies within Virgo and the blue line for those in the field.}
\label{fig:den_type}
\end{figure}

In Fig. \ref{fig:den_type} we show the density contrast parameter plotted against the morphological type (T) of each galaxy. There is quite a lot of scatter, but if we bin the data into the morphological types as defined in Table \ref{table:numbers_morph} we find a morphology density relation with the early type galaxies on average residing in the regions of highest density (Fig. \ref{fig:den_type}, red line). The Virgo cluster is obviously a very dominant feature in our data set and whether our density contrast parameter ($\delta$) measures an equivalent thing in the cluster as it does outside is not clear (there must be considerable confusion along the line of sight and in velocity). However, with this caveat in mind we have also distinguished galaxies by their cluster membership (within the virial radius). The Virgo cluster data most clearly shows a density morphology relation (Fig.  \ref{fig:den_type}), which is not apparent for galaxies in the field. At face value morphology is a consequence of being in more or less dense regions of the cluster, while density in the field has little effect on what you are. The cluster density morphology relation is of course well known and was alluded to earlier (Oemler 1974, Dressler 1980).

There is considerable scatter in the data shown in Fig. \ref{fig:den_type}, with some galaxies having quite high values of $\delta$. For example $\delta=15$ corresponds to a surface density of fourteen galaxies per Mpc$^{2}$. There are a number of galaxies in the sample that are nearby (of order 1 Mpc), leading to large angular size detection apertures.  Many also have large "peculiar" velocities that would place them at greater distances if due to the Hubble flow, for example in the Virgo cluster. This can lead to a large number of plausible neighbours over a large area of sky. However, mean values of $\delta$ are consistent (Fig. \ref{fig:den_type}). For example the approximately constant mean value of $\delta = 3$ for the field corresponds to four galaxies per Mpc$^{2}$ roughly as expected for the Local Group (as described above). Early types in Virgo reside in regions where the surface density is some three times higher than this. The measured surface density of Virgo and field late types is the same.

There is clearly a difference in the distribution of the density contrast parameter of field and cluster galaxies - the cluster shows a clear change in the morphological mix with density, the field does not (Fig. \ref{fig:den_type}). 52\% of the field sample have a density contrast of $\le$3, while this is 36\% for the Virgo sample. A density contrast of 3 corresponds to $\approx$4 galaxies per Mpc$^{2}$ and to the mean value found for all morphological types in the field (Fig. \ref{fig:den_type}).  In what follows we will use this surface density as an indicator of where we might expect galaxies to have had their properties changed due to the influence of their local environment. We will compare the properties of galaxies in the field and in Virgo, which have different morphologies and that reside in regions where the density contrast is less than or greater than $\delta=3$. 

\subsubsection{Chemical evolution}
\begin{figure}
\centering
\includegraphics[scale=0.25]{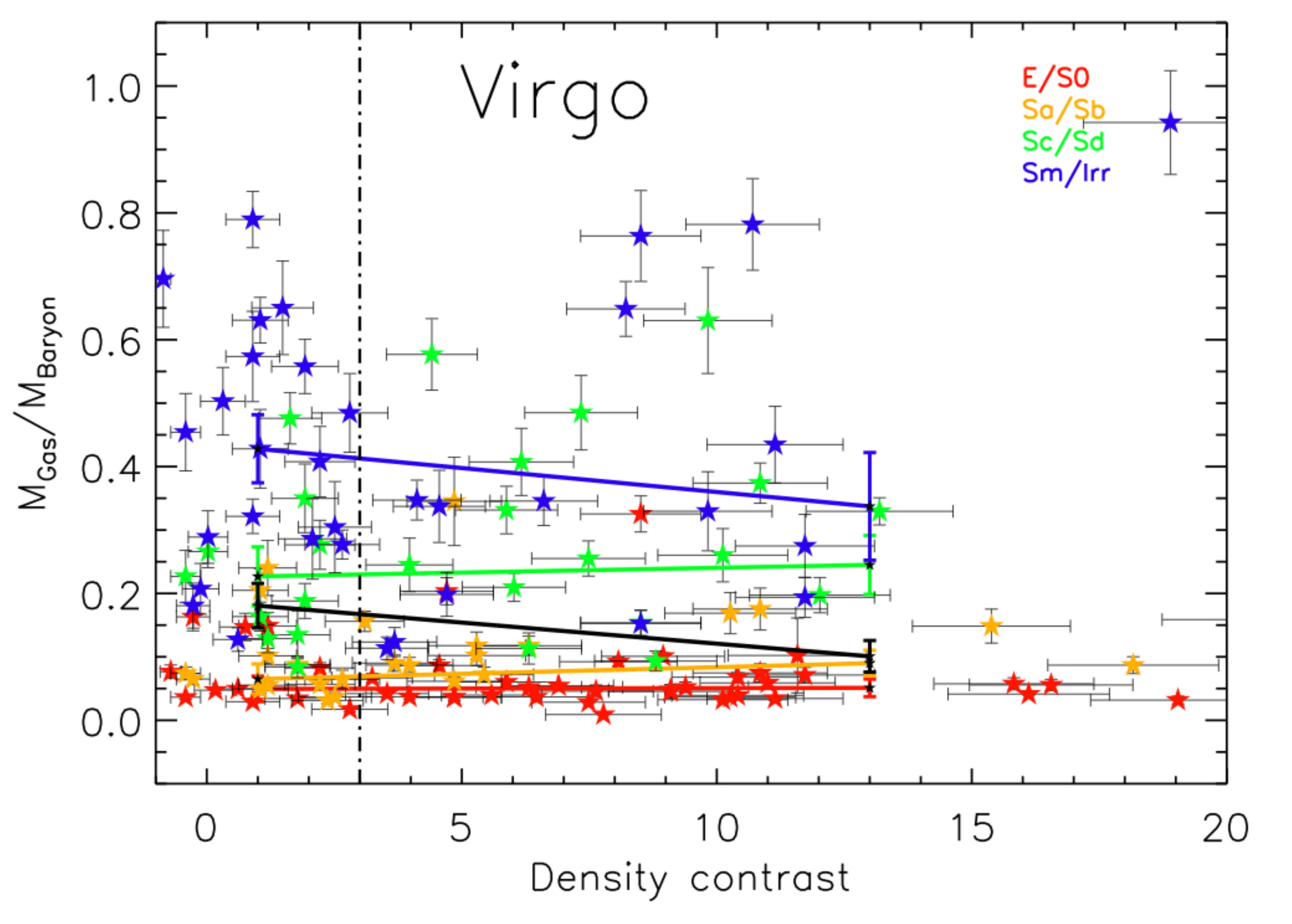}
\includegraphics[scale=0.25]{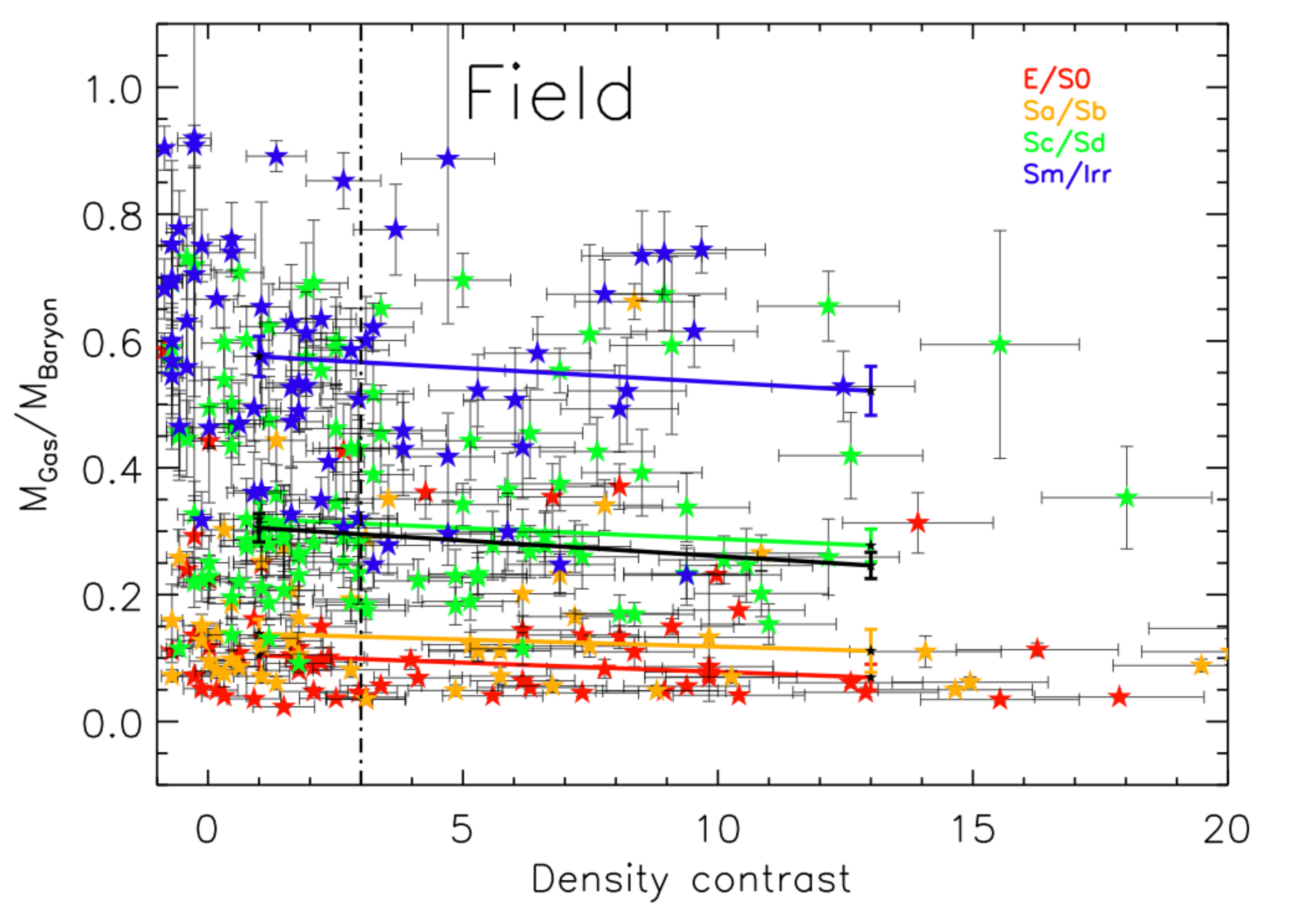}
\vspace{0.0cm}
\caption{The relationship between the gas fraction ($f=M_{Gas}/M_{Baryon}$) and the density contrast parameter ($\delta$) - Virgo (top) and Field (bottom). The data are separated into different morphological types. The coloured lines join the median values of $M_{Gas}/M_{Baryon}$ for density contrasts less than and greater than 3.0 ($\approx$4 galaxies per Mpc$^{2}$, dot-dash line). The solid black line is for the whole sample ignoring morphological type.}
\label{fig:gas_bary_ratio_density}
\end{figure}

We will use the gas fraction ($f=M_{Gas}/M_{Baryon}$) as a measure of  how chemically evolved a galaxy is i.e. how far it is along the path from total gas to total stars. In Fig. \ref{fig:gas_bary_ratio_density} we show how the gas fraction ($M_{Gas}/M_{Baryon}$) relates to the local galaxy density. Fig. \ref{fig:gas_bary_ratio_density} clearly demonstrates that the gas fractions of field galaxies, irrespective of morphological type, are higher than in the cluster - each coloured line in Fig. \ref{fig:gas_bary_ratio_density} (Virgo) is displaced to higher values in Fig. \ref{fig:gas_bary_ratio_density} (Field). So, gas fractions are higher in the field than in the cluster irrespective of differences in the morphological mix, though the effect is small for earlier type galaxies. Cluster galaxies are either more "evolved" or they have lost gas in some other way. Given our previous comments on the reasonably good fit to a closed box chemical evolution model we again suggest that they are more "evolved". This seems to be as a result of being in the cluster and is not a density effect, as Fig. \ref{fig:gas_bary_ratio_density} clearly shows that there is no measurable change in gas fraction with density contrast. This is also intriguing because it is the later types that occupy the periphery of the cluster (Fig. \ref{fig:fig_1}) where one might expect the smallest environmental impact.
Clearly belonging to a cluster has a much stronger effect on a galaxy's current gas fraction (chemical evolution) than being in other dense environments. 

\subsubsection{The specific star formation rate}
\begin{figure}
\centering
\includegraphics[scale=0.33]{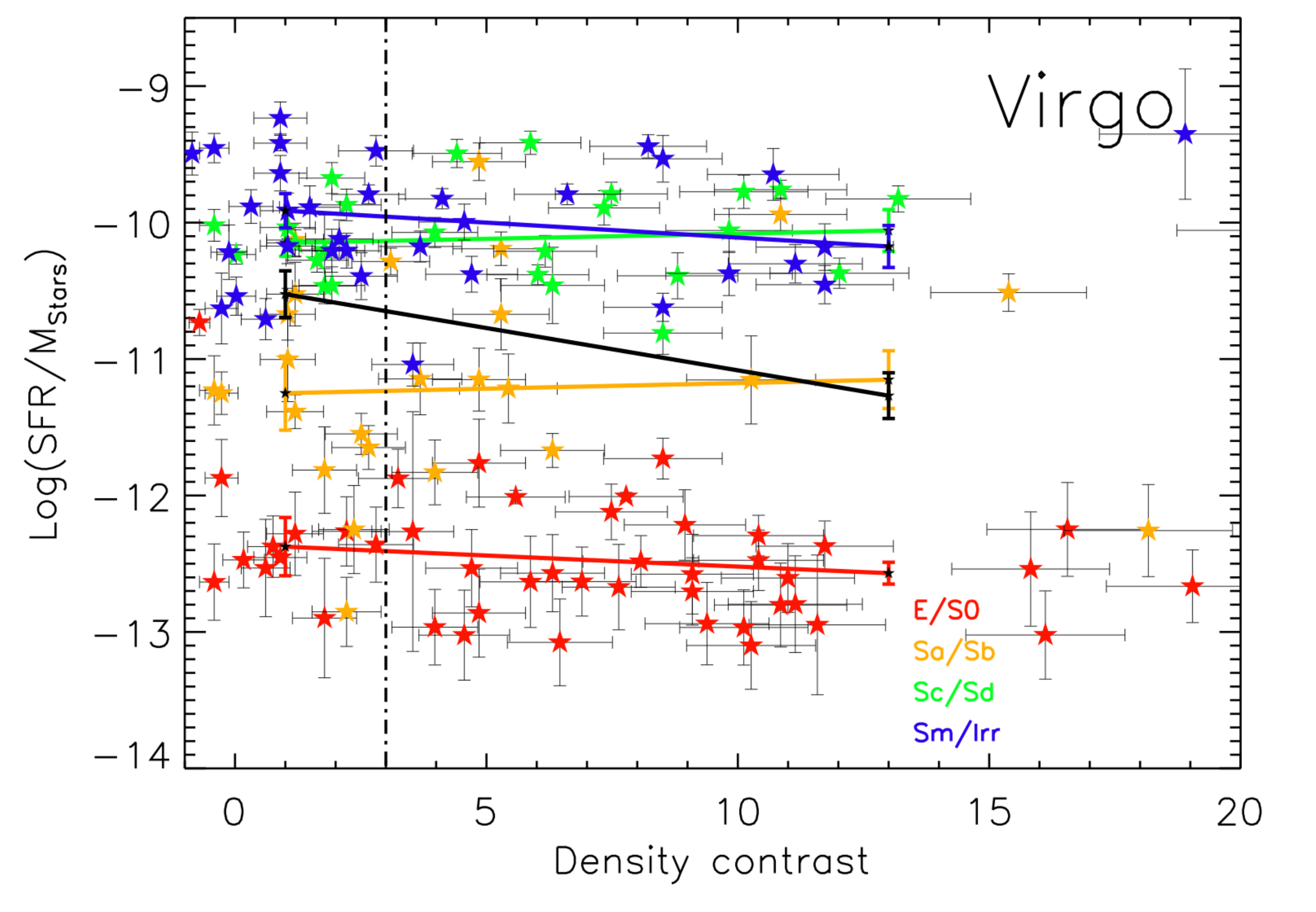}
\includegraphics[scale=0.33]{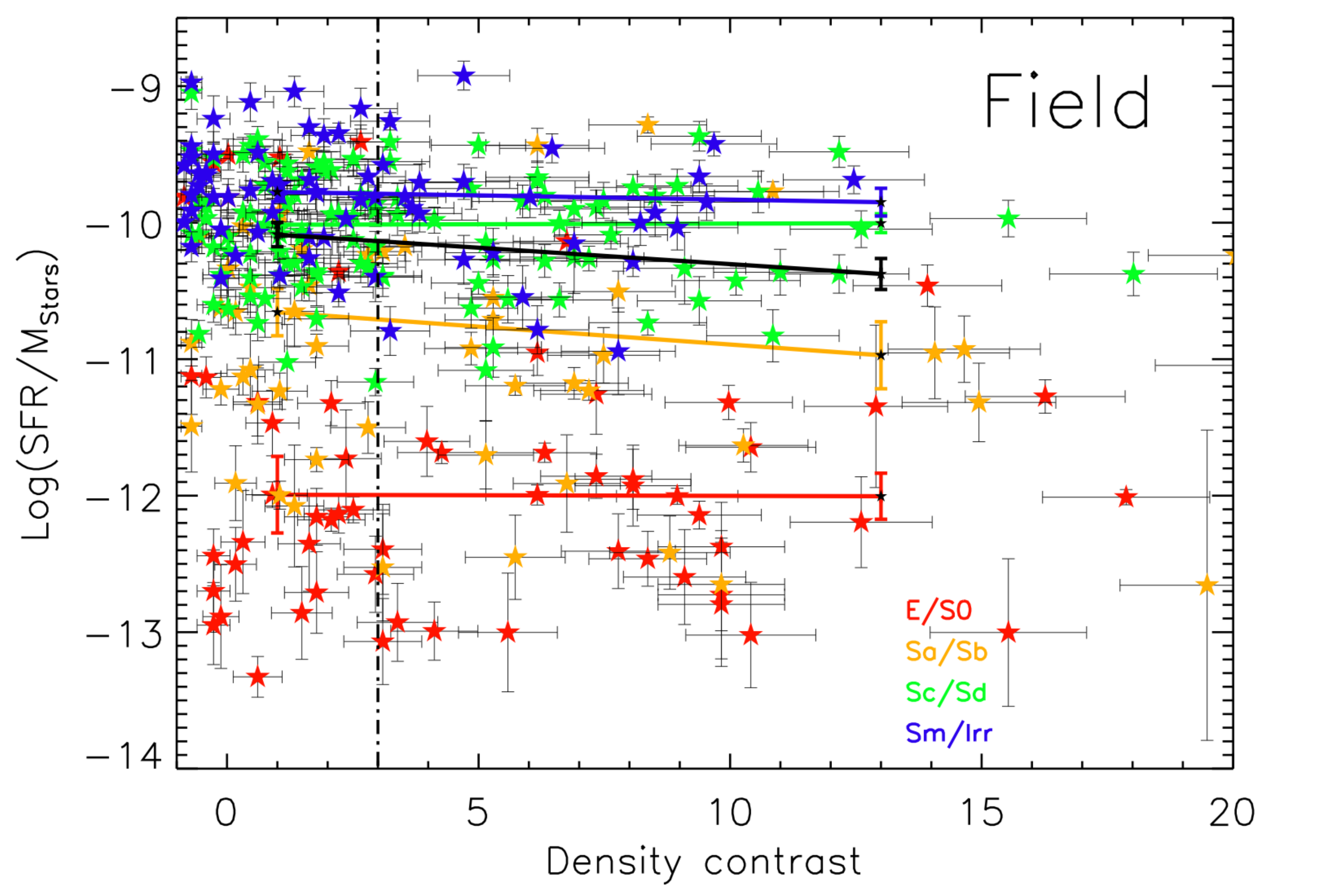}
\vspace{-0.36cm}
\caption{The relationship between the $SSFR$ ($SFR/M_{Stars}$) and the density contrast parameter ($\delta$) - Virgo (top) and Field (bottom). The data are separated by colour into different morphological types. The coloured lines join the median values of $SSFR$ for density contrasts less than and greater than 3.0 ($\approx$4 galaxies per Mpc$^{2}$, dot-dash line). The solid black line is for the whole sample ignoring morphological type.}
\label{fig:sfr_star_ratio_density}
\end{figure}

In Fig. \ref{fig:sfr_star_ratio_density} we show the SSFR ($SFR/M_{Stars}$) plotted against the density contrast. Again there is no evidence that the SSFR is changing with galaxy density for any of the morphological types whether within or outside of Virgo. Within the bounds of our SF history model a constant value of SSFR arises from common values of $t$ and $\tau$ and so the lines drawn on Fig. \ref{fig:sfr_star_ratio_density} indicate common SF histories irrespective of local galaxy density. The black line on Fig. \ref{fig:sfr_star_ratio_density} (top) does show  a decline in SSFR with density, but this is clearly produced by the change in the morphological mix, there being relatively more early type galaxies in high densities in Virgo. This is a further illustration of why a proper understanding of the morphological mix of a galaxy sample is so important before making comparisons with other samples.

\subsubsection{The star formation efficiency}
\begin{figure}
\centering
\includegraphics[scale=0.33]{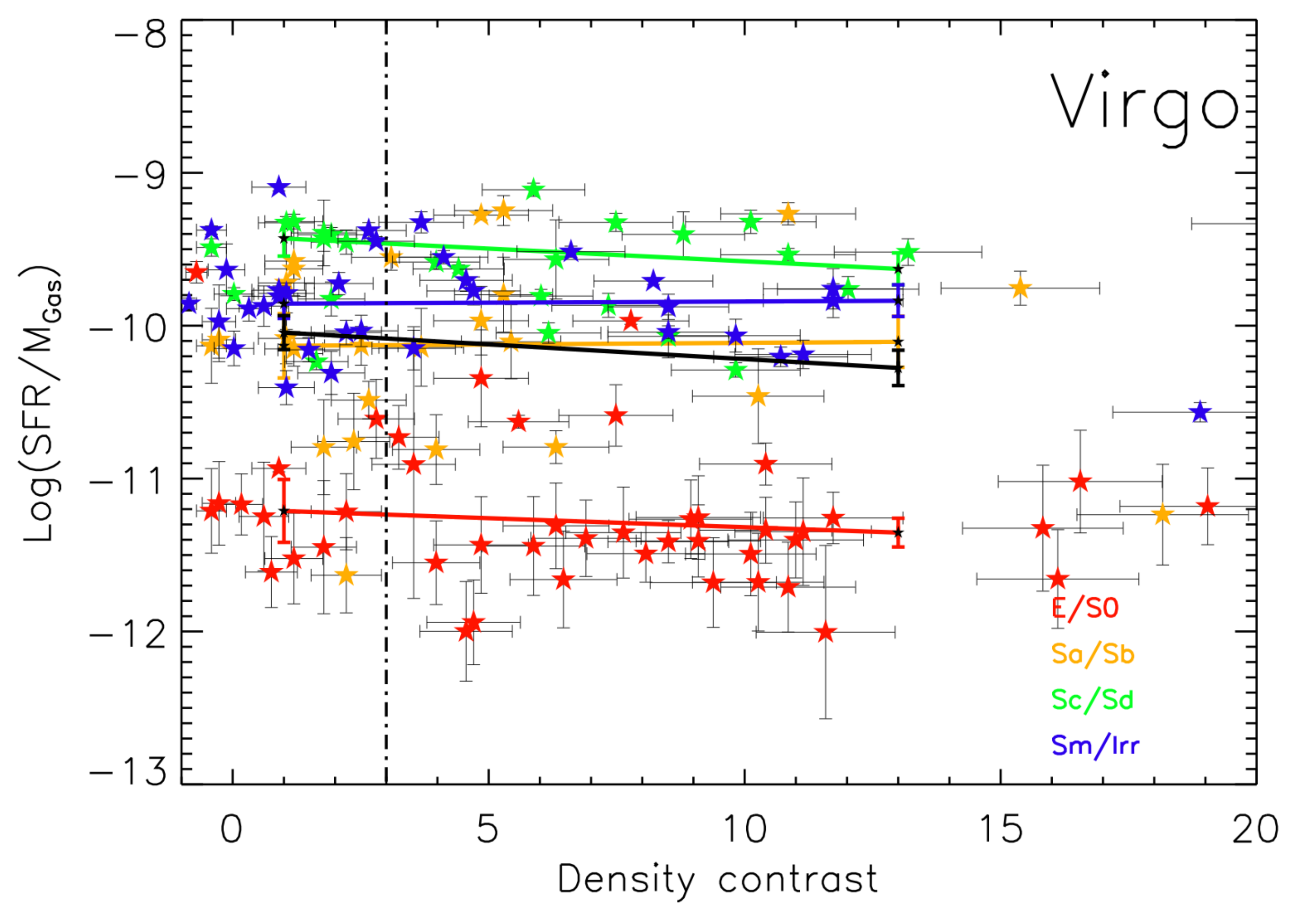}
\includegraphics[scale=0.33]{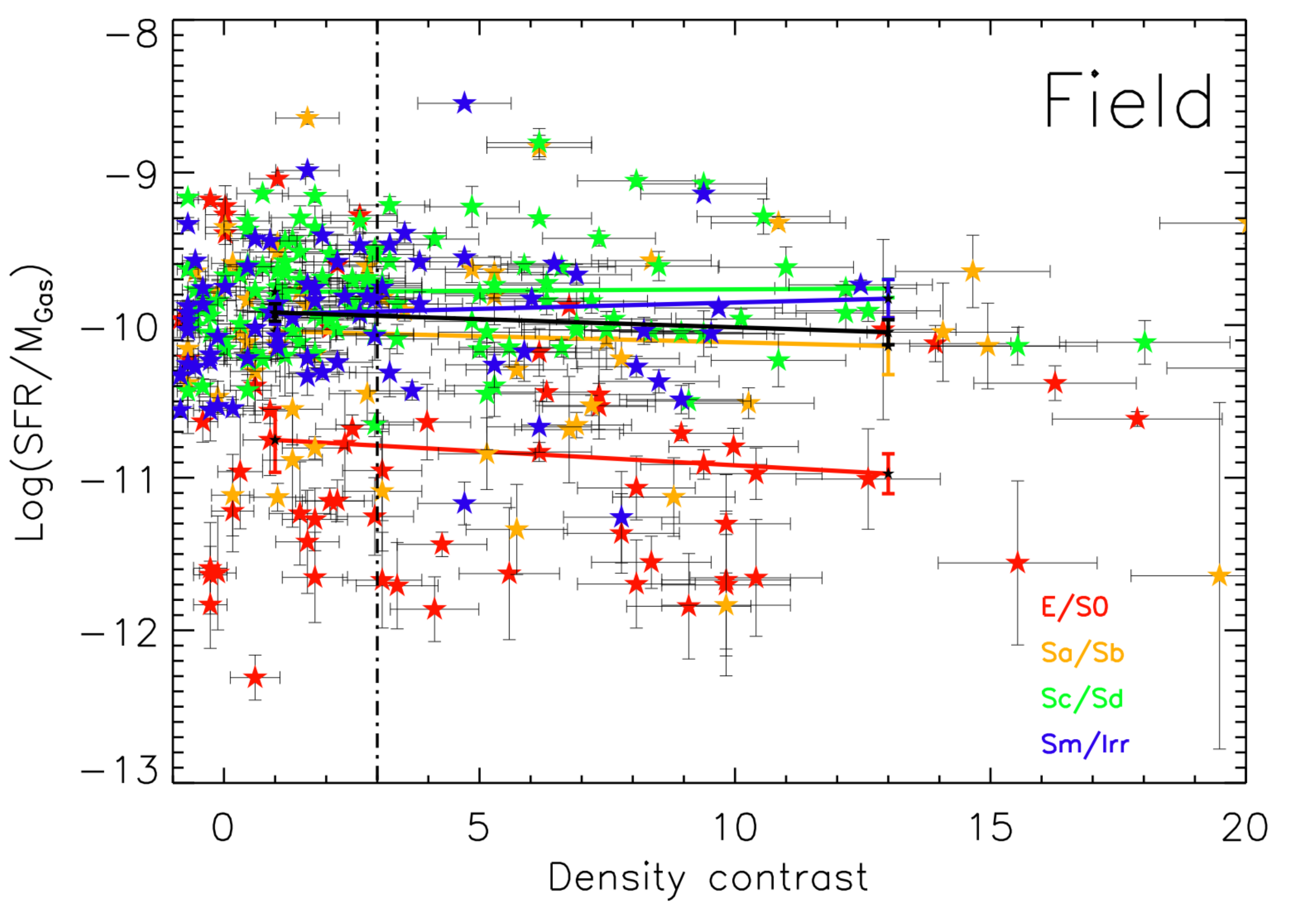}
\vspace{-0.36cm}
\caption{The relationship between the star formation efficiency ($SFR/M_{Gas}$) and the density contrast parameter ($\delta$) - Virgo (top) and Field (bottom). The data are separated by colour into different morphological types. The coloured lines join the median values of $SSFR$ for density contrasts less than and greater than 3.0 ($\approx$4 galaxies per Mpc$^{2}$, dot-dash line). The solid black line is for the whole sample ignoring morphological type.}
\label{fig:sfr_gas_ratio_density}
\end{figure}

With regard to the SFE ($SFR/M_{Gas}$) we again find no change with density contrast (Fig. \ref{fig:sfr_gas_ratio_density}). Again within the bounds of our simple model this implies that galaxies of the same morphology in different environments can only have small differences in their star formation histories. As before morphology is much more important with regard to SFE than any environmental affects. It is intriguing that although field galaxies have larger gas fractions (Fig. \ref{fig:gas_bary_ratio_density}) they do not seem to have any differences in their SFEs when compared to cluster galaxies. However, looking at Fig. \ref{fig:sf_eff} we see that the predicted relationship (blue dot and dashed lines) between gas fraction and SFE is actually quite flat. Over a good fraction of a galaxy's life ($0.1<f<1.0$, for example) its SFE is predicted to change by only a factor of 10. Thus small differences in the gas fraction of Virgo galaxies are reflected in consequentially small changes in the SFE. For example from Fig. \ref{fig:gas_bary_ratio_density} we see that the gas fraction of late type (Sm/Irr) galaxies changes from about 0.4 to 0.6 between Virgo and the field. From the star formation history model (section 4.1.4) this leads to a predicted change in the value of $\log{SFE}$ of $<0.2$, small compared to the error bars and scatter in the data shown in Fig. \ref{fig:sfr_gas_ratio_density}.

\subsubsection{The stars-to-dust mass ratio}
\begin{figure}
\centering
\includegraphics[scale=0.33]{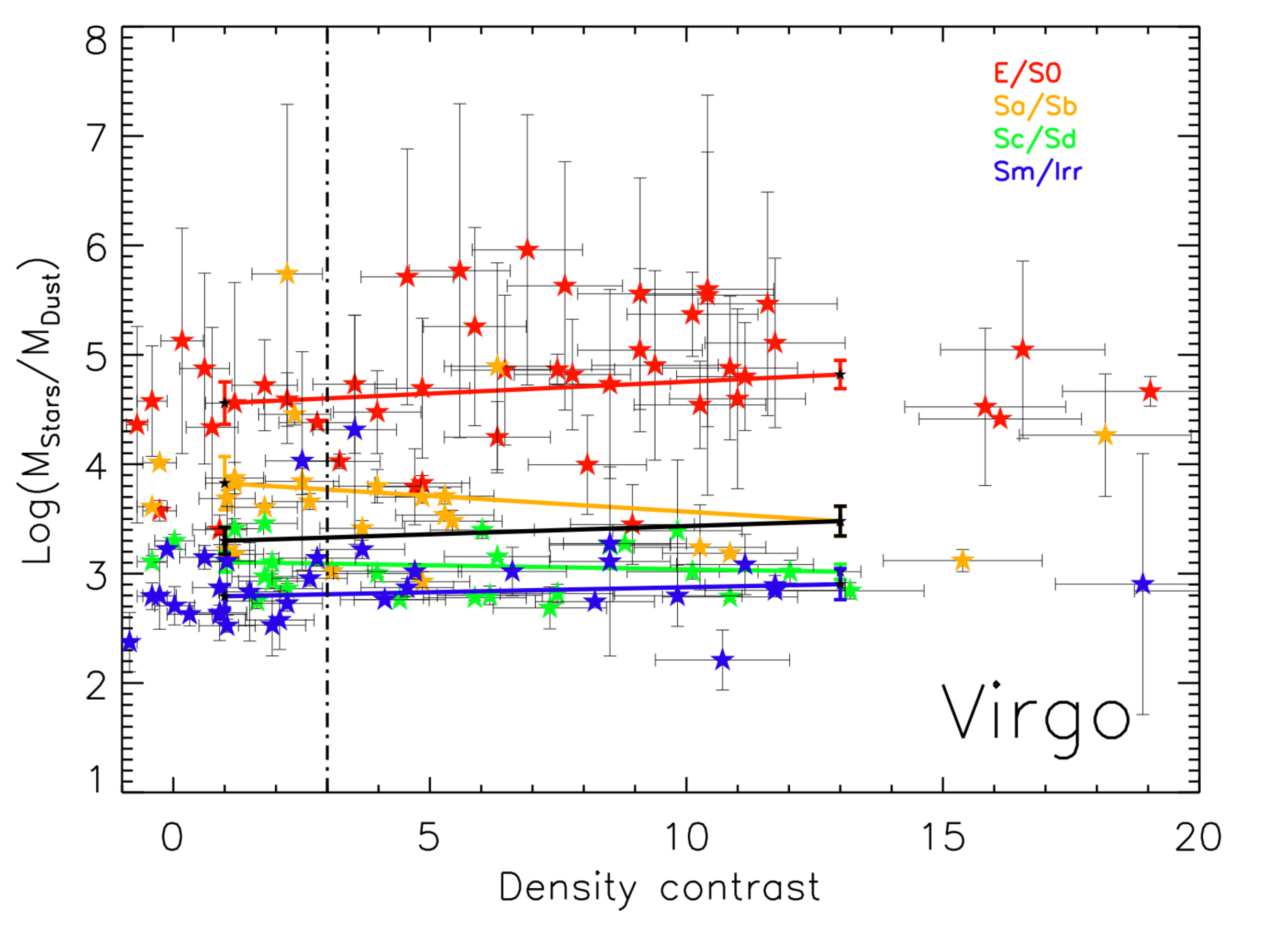}
\includegraphics[scale=0.33]{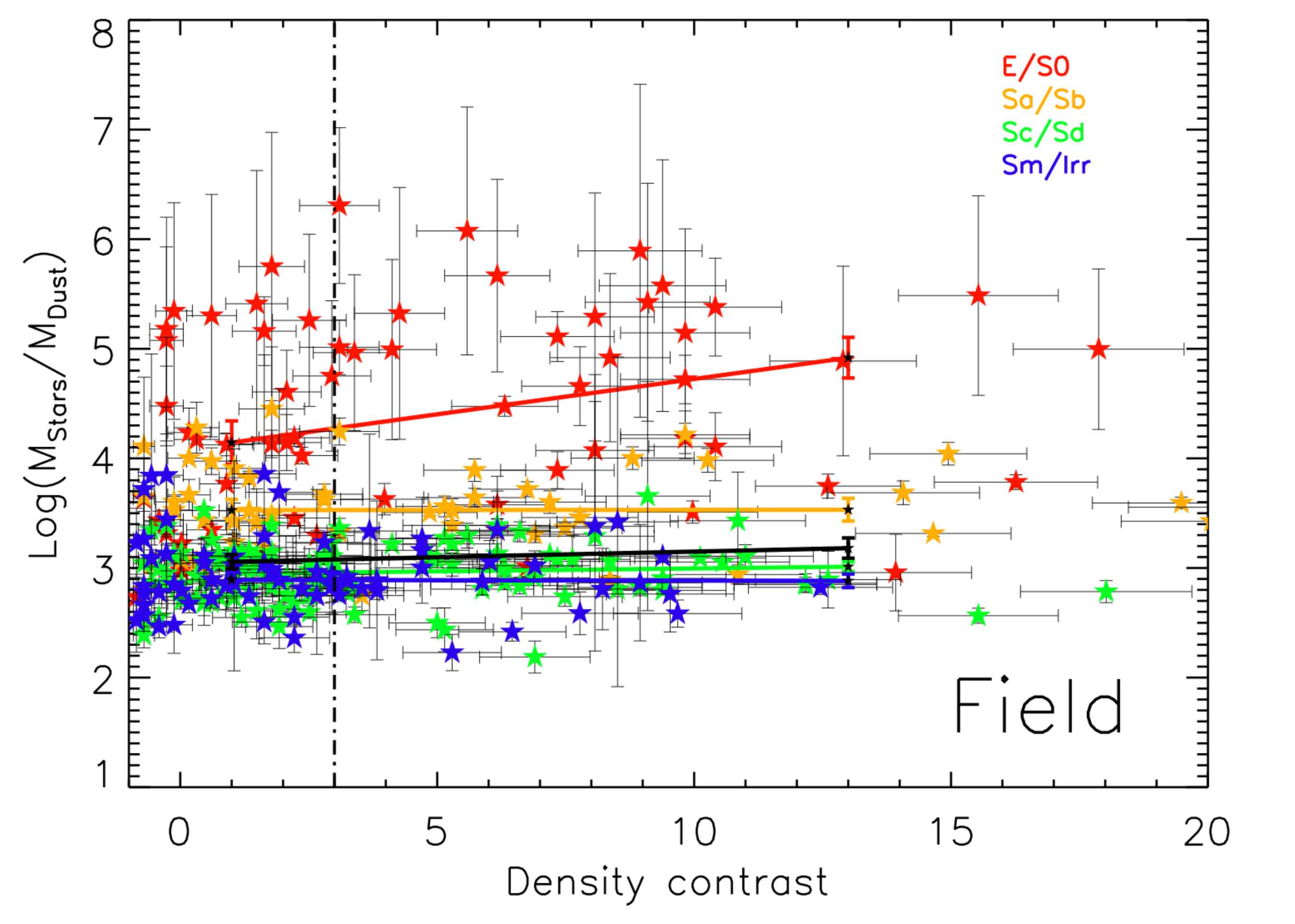}
\vspace{-0.36cm}
\caption{The relationship between the ratio $M_{Stars}/M_{Dust}$ and the density contrast parameter ($\delta$) - Virgo (top) and Field (bottom). The data are separated by colour into different morphological types. The coloured lines join the median values of $M_{Gas}/M_{Dust}$ for density contrasts less than and greater than 3.0 ($\approx$4 galaxies per Mpc$^{2}$, dot-dash line). The solid black line is for the whole sample ignoring morphological type.}
\label{fig:star_dust_ratio_density}
\end{figure}

In Fig. \ref{fig:star_dust_ratio_density} we plot the $M_{Star}/M_{Dust}$ ratio of each galaxy against its density contrast. Again there is little change in $M_{Star}/M_{Dust}$ with density contrast and morphology is much more important than the local environment. The one possible exception is the higher value of $M_{Star}/M_{Dust}$ for field, and possibly Virgo, early type (E/S0) galaxies at higher density contrasts. 

Considering our previously derived relationship  in section 4.1.5 between $M_{Star}/M_{Dust}$ and $p$, $\eta$, $t$ and $\tau$. The simplest interpretation of Fig. \ref{fig:star_dust_ratio_density} is that the stellar yield ($p$), fraction of metals in the dust ($\eta$), time of formation ($t$) and the time after formation that the SFR peaked ($\tau$) are pretty much the same for galaxies of the same morphology in different environments.

\subsubsection{The gas-to-dust mass ratio}
\begin{figure}
\centering
\includegraphics[scale=0.33]{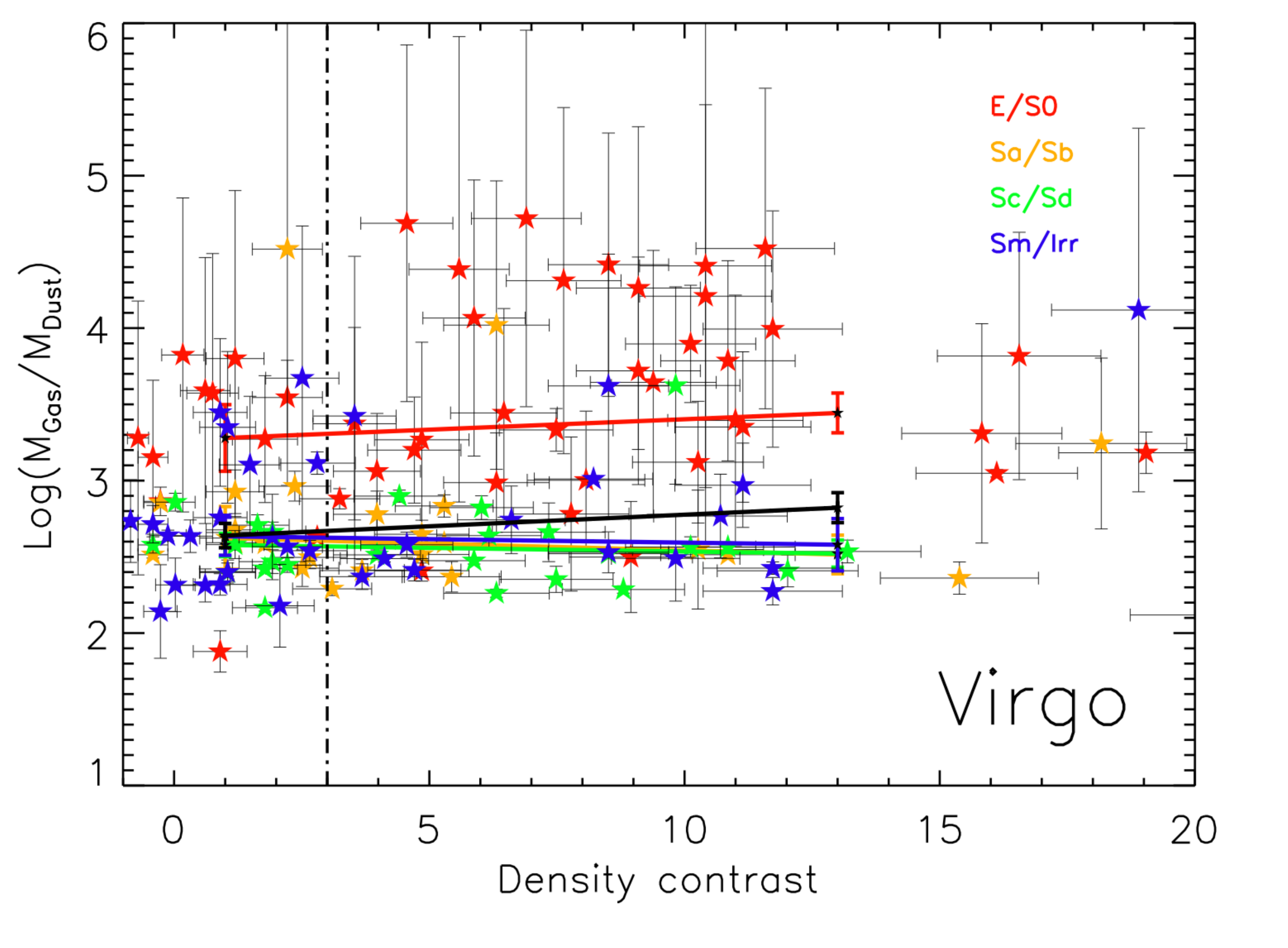}
\includegraphics[scale=0.33]{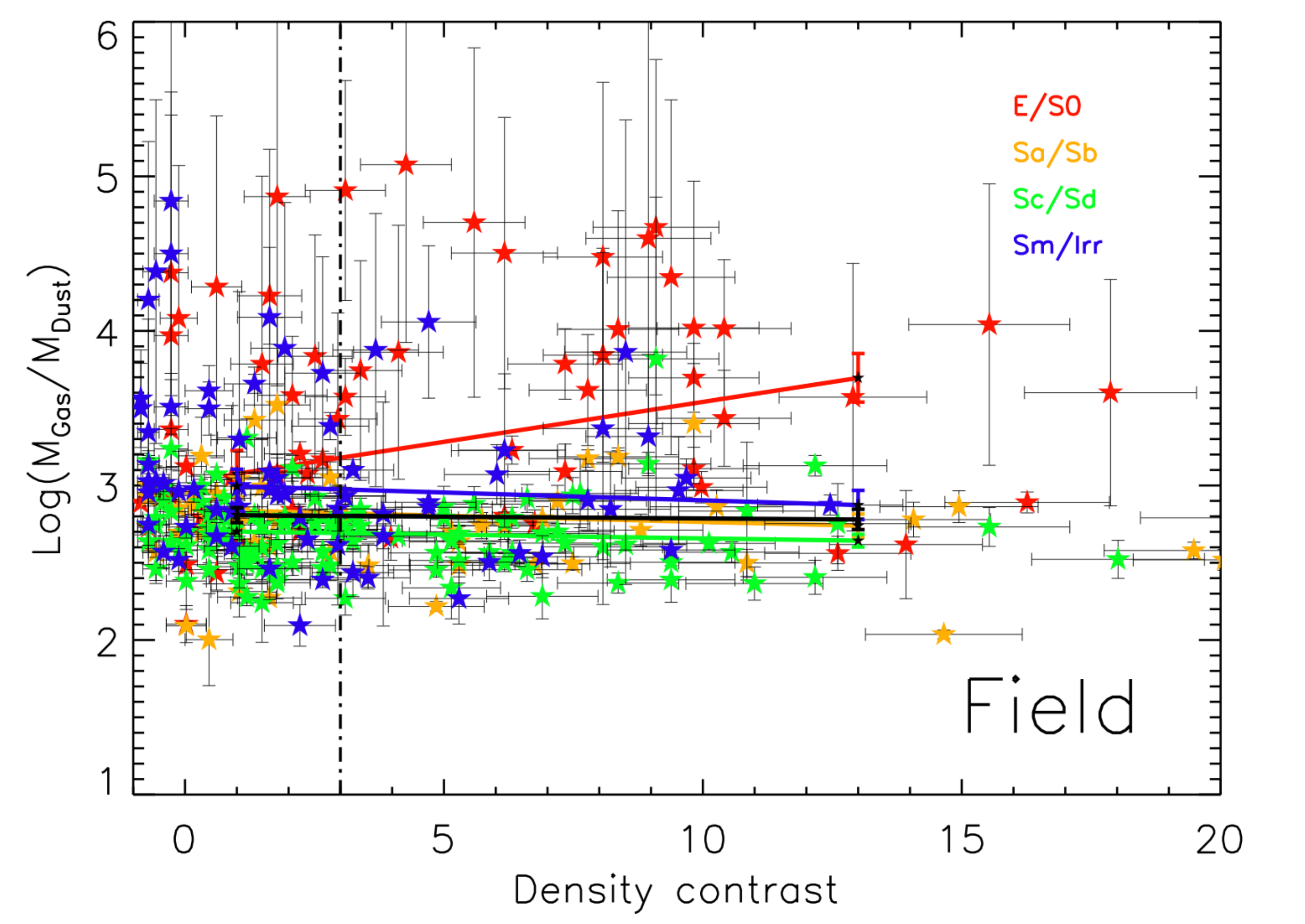}
\vspace{0.0cm}
\caption{The relationship between the ratio $M_{Gas}/M_{Dust}$ and the density contrast parameter ($\delta$) - Virgo (top) and Field (bottom). The data are separated by colour into different morphological types. The coloured lines join the median values of $M_{Gas}/M_{Dust}$ for density contrasts less than and greater than 3.0 ($\approx$ 4 galaxies per Mpc$^{2}$, dot-dash line). The solid black line is for the whole sample ignoring morphological type.}
\label{fig:gas_dust_ratio_density}
\end{figure}

In Fig. \ref{fig:gas_dust_ratio_density} we consider the ratio $M_{Gas}/M_{Dust}$ in relation to galaxy density.  With the exception of the early type (E/S0) galaxies we find no change in $M_{Gas}/M_{Dust}$ with density contrast. There is a higher value of $M_{Gas}/M_{Dust}$ for field, and possibly Virgo, early type (E/S0) galaxies at higher density contrasts. We have no explanation of why this is so, but we note that the early type galaxies have been almost impossible to model in every plot we have made! Particularly in Fig. \ref{fig:sf_hist} we showed that within the bounds of our simple model, values of $M_{Star}/M_{Dust}$ and $M_{Gas}/M_{Dust}$ are inconsistent with each other for early type galaxies. We shall return to this in section 4.2.7 below.

\subsubsection{The dust temperature}
\begin{figure}
\centering
\includegraphics[scale=0.33]{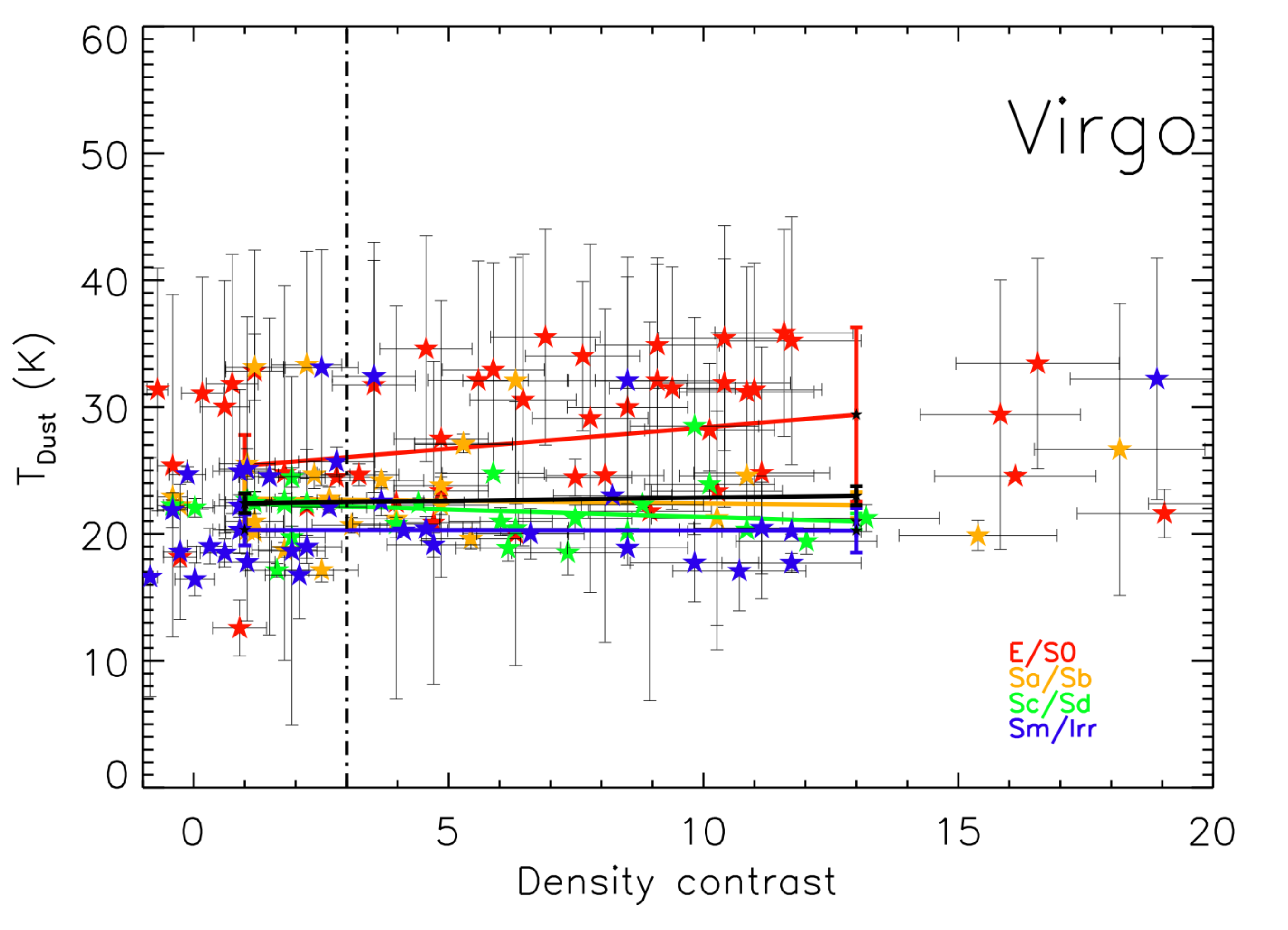}
\includegraphics[scale=0.33]{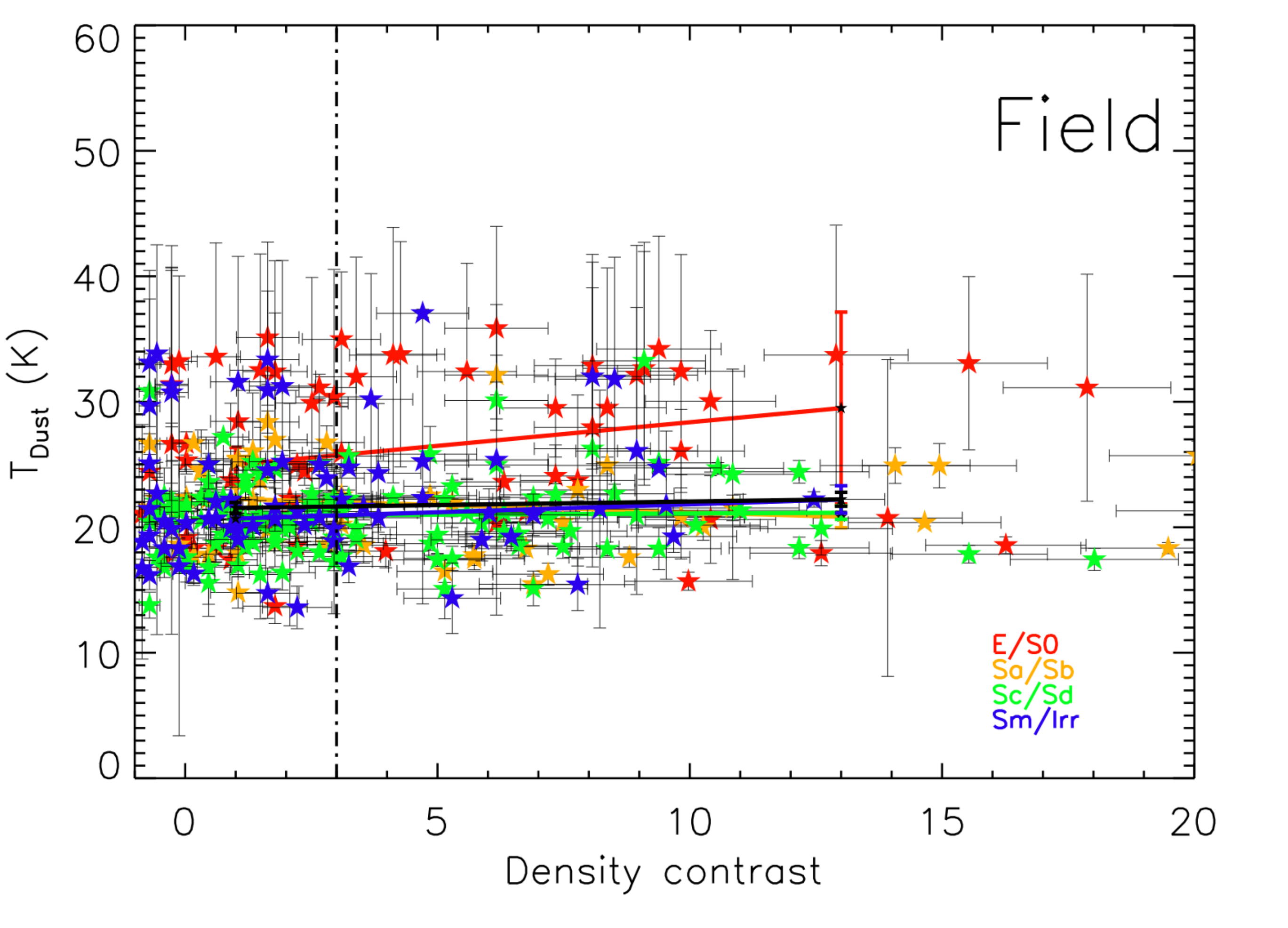}
\vspace{-0.36cm}
\caption{The relationship between dust temperature (K) and the density contrast parameter ($\delta$) - Virgo (top) and Field (bottom). The data is separated into different morphological types with the dashed lines being a linear fit to each type. The black dot-dashed line indicates a galaxy surface density $\approx$4 galaxies per Mpc$^{2}$.}
\label{fig:temp_density}
\end{figure}

For completeness we show in Fig. \ref{fig:temp_density} the relationship between dust temperature and density contrast. It is plausible that galaxies could have, for example, warmer dust temperatures as a result of interactions and mergers in denser environments or even additional heating via an x-ray gas that they may reside within. However,   Fig. \ref{fig:temp_density} shows again that there is no real evidence for a change of dust temperature with galaxy density. The possible exception is again early type galaxies (E/S0) in the cluster where there is a noticeable increase in temperature for galaxies in denser environments (possibly heating by x-ray emitting hot gas) though the error on the median values is large.

\subsubsection{Comments on early type galaxies}
In almost all of the above the late type galaxies have corresponded very well with the expectations of the combined simple SF history and chemical evolution model we have used - the later the type the better it all seems to work. In contrast the early type galaxies (E/S0) have almost always failed to comply. The issues are:
\begin{enumerate}
\item They do not lie on the "main sequence" and probably do not lie on any sequence at all (Fig. 4).
\item Unlike the late types they do not lie on a line that provides consistent values for the SF history parameters $t$ and $\tau$ (Fig. 5) - they are predicted to be too old.
\item Unlike the linear relation for late type galaxies there is large scatter in values of dust mass at a given stellar and gas mass for early types (Figs. 6 and 7).
\item Within the bounds of the model, values of $t$ and $\tau$ derived from $M_{Star}/M_{Dust}$ and $M_{Gas}/M_{Dust}$ are inconsistent.
\item Dust in late type galaxies is clearly heated by SF. In early type galaxies it is something else (Fig. 9).
\item There is no evidence for any differences in the properties of late type galaxies changing with environment as measured by the density contrast, though there are differences in gas fraction between cluster and field (Figs 11-16). However, there are "hints" that for early type galaxies $M_{Star}/M_{Dust}$, $M_{Gas}/M_{Dust}$ and dust temperature are all higher in more dense environments.
\end{enumerate}

Clearly the simple model is an inadequate interpretation of the derived physical parameters of the early type galaxies considered here. An obvious solution is that while late types are plausibly modelled  as closed box  systems, early type galaxies have probably been subject to more complex processes, that ingress or expel material as the galaxy evolves. However, there is little evidence that these processes are related to the environment the galaxies now find themselves within.

\section{Summary}
We have used a sub-set of the DustPedia galaxy sample along with data from the SDSS spectroscopic survey to investigate how galaxies may be affected by their local environment. For each DustPedia galaxy we have distinguished it via its membership or otherwise of the Virgo cluster (cluster or field) and have calculated a density contrast parameter using the proximity of SDSS galaxies. We also have galaxy morphologies and so are able to look for changes in dust properties dependent on density contrast, cluster membership and morphology. For each galaxy we have used the CIGALE SED fitting package with the THEMIS dust model to derive star formation rates, stellar and dust masses and dust temperatures. Atomic gas masses have been obtained separately from the literature. 

Using the above data we have considered :
\begin{enumerate}
\item Chemical evolution within the bounds of a simple closed box model and deviations from it that indicate more complicated evolution.
\item The SSFR and how it relates to different SF histories.
\item The SFE and how it relates to galaxies with different gas fractions.
\item The stars to dust and gas to dust mass ratios and whether they predict a consistent picture (using the simple model) of how a galaxy evolves.
\item The relationship between SFR, dust mass and dust temparature and what this tells us about the dominant dust heating sources.
\end{enumerate}

Our inferences from the data are:
\begin{itemize}
\item The average spectral energy distributions of galaxies of different morphologies in the cluster and field are, with two possible exceptions, the same. Firstly, at  low statistical significance early type (E/S0) cluster galaxies may have more far infrared emission for the same stellar emission. Secondly, late type galaxies in the field have more UV/blue emission than cluster galaxies. The latter is more convincing as it is apparent and progressively stronger as you move from earlier (Sa/Sb) to later (Sm/Irr) types.
\item Both cluster and field samples give a reasonable fit a simple closed box chemical evolution model. The normalisation of the model fit to the data implies an effective yield that is the same for both cluster and field samples even though we do measure higher gas fractions for the field sample. With the same effective yield there is no evidence of gas stripping or infall having a more important effect on cluster compared to field galaxies. From the chemical evolution model alone we would infer that the cluster sample is just more evolved in the sense of its progression from a gas fraction of one to zero.
\item Our measured  SSFRs as a function of stellar mass in both the cluster and field samples are consistent with each other and with previous work. There is no evidence that the "main sequence", "green valley" or sequence of "non-star forming" galaxies are different between the cluster and field. Within the bounds of our simple model this implies a common star formation history. Morphologically galaxies are well separated into types; later than Sc that lie on the main sequence, Sa/Sb that tend to lie in the green valley and E/S0 galaxies that lie on the "sequence" of non-star forming galaxies, which is somewhat less well defined. As with the inference from the chemical evolution model there is a plausible evolutionary link, in this case in relation to SF history, that explains the derived values of SSFR. The simple model does not provide an explanation of any stellar mass dependence on SSFR or how and why morphological transformation seems to change in sync with chemical evolution and SSFR.
\item Assuming the simple closed box model we infer that the SFE of a galaxy peaks when about 75\% of the gas has been consumed independent of its environment. We measure consistent SFEs for cluster and field galaxies, which within the bounds of our simple model, again implies a common star formation history.  
\item  Our data is consistent with an almost constant value of $M_{Star}/M_{Dust} \approx 10^{3}$ for types later than Sc irrespective of whether they are in the cluster or field. Within the bounds of our simple model this value of $M_{Star}/M_{Dust}$ implies that the star formation peak of these galaxies occurs consistently about half way through their lives at $t/\tau \approx 2$. The wide ranging values of $M_{Star}/M_{Dust}$ for the early types (particularly E/S0) is difficult to explain within the bounds of the simple model and may require external sources of dust, for example through mergers. Our simple model predicts a minimum value of $\log{M_{Star}/M_{Dust}}=2.5$.
\item For late type galaxies (Sc/Sd, Sm/Irr) we find no difference in the value of $M_{Gas}/M_{Dust} \approx 10^{2.7}$ between cluster and field. Early types (E/S0) have larger values of $M_{Gas}/M_{Dust}$ and consistent values between cluster and field, but again there is considerable scatter in the data with there being a large range of dust mass at a given gas mass. This is again difficult to reconcile within the bounds of our simple model.
\item Mean values of $M_{Stars}/M_{Dust}$ and $M_{Gas}/M_{Dust}$ for different morphologies can be used to check for consistent values of the star formation history parameters  $t$ and $\tau$. Late type galaxies fit into a consistent picture of closed box chemical evolution and a star formation rate that rises to a peak and then declines as specified by our model. Mean values of $M_{Stars}/M_{Dust}$ and $M_{Gas}/M_{Dust}$ for early type galaxies produce inconsistent star formation histories and so will require additions to our model interpretations, though we stress that we find no evidence that their evolution has been altered by their presence or other wise in the cluster environment.
\item For morphological types later than Sc dust heating is consistent with it being due to recently formed stars. For early types heating is primarily not associated with SF - possibilities are a X-ray gas, energetic electrons or the general stellar radiation field. Thus the most important dust heating modes are morphology dependent. This conclusion seems to be independent of the environment, whether cluster or field. 
\item Differences in dust temperature between galaxies appear to be morphologically type dependent and not environmentally dependent.
\item In addition to the above comparisons between cluster and field galaxies we have also compared galaxy properties with the local number density of galaxies. We have used SDSS data to define a density contrast parameter for each DustPedia galaxy in our sample. We have then compared the properties ($\log{M_{Gas}/M_{Baryon}}$, $\log{SFR/M_{Stars}}$, $\log{SFR/M_{Gas}}$, $\log{M_{Star}/M_{Dust}}$, $\log{M_{Gas}/M_{Dust}}$ and T$_{Dust}$) of these galaxies for those galaxies that reside within surface densities of less than and greater than 4 galaxies per Mpc$^{2}$. Essentially we find no measurable differences in the above properties of galaxies of the same morphology in different density environments. It is morphology that defines the properties of these galaxies and not their environment. This is most clearly illustrated in the field where we find no morphology density relation and so the local environment has not affected the morphological mix of galaxies. Thus their properties arise purely as a result of their environmentally independently derived morphology.
\end{itemize}

Much of the above supports a previous conclusion to the same effect by Park et al. (2007) who considered the optical properties of galaxies. They rather succinctly said that "When morphology and luminosity are fixed, other physical properties, such as colour, colour gradient, concentration, size, velocity dispersion, and star formation rate, are nearly independent of local density, without any break or feature." We can extend this conclusion to both the dust and gas properties of galaxies. Clearly differences in the galaxy properties we have measured vary to a much greater extent as a function of morphology than as a function of local density (see also Bait et al. 2017). 

Although we have not considered explicitly variations in galaxy properties as a function of stellar luminosity, we find no obvious differences in the measured properties of galaxies, with $M_{Stars}$ less than or greater than a value that corresponds to $M^{*} \approx 10^{10}$ M$_{\odot}$, as suggested by Robotham et al. (2014). Thus, there is no indication in our data that galaxies fainter than M$^{*}$ predominately grow their stellar mass via in-situ star formation, while those that are brighter primarily grow through mergers. However, at higher stellar masses the morphological mix is increasingly dominated by early type galaxies, which we have shown to have much more complicated SF histories.

We conclude that galaxies clearly have properties that are morphological type dependent, but there is very little evidence that the local environment has had a significant influence on galaxies of the same morphological type. It appears that it is primarily morphology, how and whenever this is laid down, and consistent internal physical processes that determine these properties of galaxies in the DustPedia sample. The key seems to be not how galaxies have reacted and changed due to their environment, but how they achieved their morphology in the first place.

\vspace{0.5cm}
\noindent
\large
{\bf Acknowledgements} \\
\normalsize
DustPedia is a collaborative focused research project supported by European Union Grant 606847 awarded under the FP7 call. Further information can be found at www.dustpedia.com.

Funding for the Sloan Digital Sky Survey IV has been provided by
the Alfred P. Sloan Foundation, the U.S. Department of Energy Office of
Science, and the Participating Institutions. SDSS-IV acknowledges
support and resources from the Center for High-Performance Computing at
the University of Utah. The SDSS web site is www.sdss.org.

SDSS-IV is managed by the Astrophysical Research Consortium for the 
Participating Institutions of the SDSS Collaboration including the 
Brazilian Participation Group, the Carnegie Institution for Science, 
Carnegie Mellon University, the Chilean Participation Group, the French Participation Group, Harvard-Smithsonian Center for Astrophysics, 
Instituto de Astrof\'isica de Canarias, The Johns Hopkins University, 
Kavli Institute for the Physics and Mathematics of the Universe (IPMU) / 
University of Tokyo, Lawrence Berkeley National Laboratory, 
Leibniz Institut f\"ur Astrophysik Potsdam (AIP),  
Max-Planck-Institut f\"ur Astronomie (MPIA Heidelberg), 
Max-Planck-Institut f\"ur Astrophysik (MPA Garching), 
Max-Planck-Institut f\"ur Extraterrestrische Physik (MPE), 
National Astronomical Observatories of China, New Mexico State University, 
New York University, University of Notre Dame, 
Observat\'ario Nacional / MCTI, The Ohio State University, 
Pennsylvania State University, Shanghai Astronomical Observatory, 
United Kingdom Participation Group,
Universidad Nacional Aut\'onoma de M\'exico, University of Arizona, 
University of Colorado Boulder, University of Oxford, University of Portsmouth, 
University of Utah, University of Virginia, University of Washington, University of Wisconsin, 
Vanderbilt University, and Yale University.

We acknowledge the usage of the HyperLeda database (http://leda.univ-lyon1.fr).

This research has made use of the NASA/IPAC Extragalactic Database (NED) which is operated by the Jet Propulsion Laboratory, California Institute of Technology, under contract with the National Aeronautics and Space Administration. 

The Herschel spacecraft was designed, built, tested, and launched under a contract to ESA managed by the Herschel/Planck Project team by an industrial consortium under the overall responsibility of the prime contractor Thales Alenia Space (Cannes), and including Astrium (Friedrichshafen) responsible for the payload module and for system testing at spacecraft level, Thales Alenia Space (Turin) responsible for the service module, and Astrium (Toulouse) responsible for the telescope, with in excess of a hundred subcontractors.

PACS has been developed by a consortium of institutes led by MPE (Germany) and including UVIE (Austria); KU Leuven, CSL, IMEC (Belgium); CEA, LAM (France); MPIA (Germany); INAF-IFSI/OAA/OAP/OAT, LENS, SISSA (Italy); IAC (Spain). This development has been supported by the funding agencies BMVIT (Austria), ESA-PRODEX (Belgium), CEA/CNES (France), DLR (Germany), ASI/INAF (Italy), and CICYT/MCYT (Spain).

SPIRE has been developed by a consortium of institutes led by Cardiff University (UK) and including Univ. Lethbridge (Canada); NAOC (China); CEA, LAM (France); IFSI, Univ. Padua (Italy); IAC (Spain); Stockholm Observatory (Sweden); Imperial College London, RAL, UCL-MSSL, UKATC, Univ. Sussex (UK); and Caltech, JPL, NHSC, Univ. Colorado (USA). This development has been supported by national funding agencies: CSA (Canada); NAOC (China); CEA, CNES, CNRS (France); ASI (Italy); MCINN (Spain); SNSB (Sweden); STFC, UKSA (UK); and NASA (USA).

This publication makes use of data products from the Wide-field Infrared Survey Explorer, which is a joint project of the University of California, Los Angeles, and the Jet Propulsion Laboratory/California Institute of Technology, funded by the National Aeronautics and Space Administration.

\vspace{0.5cm}
\noindent
\large
{\bf References} \\
Ade P. et al., 2016, AA, 594, 13 \\
Bait O., Barway S., Wadadekar Y., 2017, 471, 2687 \\
Baldry I., et al., 2006, MNRAS, 373, 469 \\
Bamford S., et al., 2009, MNRAS, 393, 1324 \\
Bianchi S., et al. 2018, 1810.01208 \\
Blanton M., Eisenstein D., Hogg D., Schlegel D. and Brinkmann J., 2005, ApJ, 629, 143 \\
Boquien M., et al., 2019, AA, 622, 103 \\
Boselli A. and Gavazzi G., 2006, PASP, 118, 517 \\
Bruzual G., Charlot S., 2003, MNRAS, 344, 1000 \\
Buat V., et al., 2007, AA, 469, 19 \\
Buitrago F., Trujillo I., Conselice C. and Haubler B., 2013, MNRAS, 428, 1460 \\
Calvi R., Vulcani B., Poggianti B., Moretti A., Fritz J. and Fasano G., 2018, MNRAS, 481, 3456 \\
Calzetti D., Armus L., Bohlin R., Kinney A., Koornneef J., Storchi-Bergmann T., 2000, ApJ, 573, 682 \\
Casasola V., Bettoni D. and Galletta G., 2004, AA, 422, 941 \\
Ciesla L., et al., 2016, AA, 585, 43 \\
Clark C., et al., 2015, MNRAS, 452, 397 \\
Clark C., et al., 2018, AA, 609, 37 \\
Clemens M., et al., 2013, MNRAS, 433, 695 \\
Cortese L., et al., 2016a, 463, 170 \\
Conselice C., Chapman S. and Windhorst R., 2003, ApJ, 596, 5 \\
Conselice C., Blackburne J. and Papovich C., 2005, ApJ, 620, 564 \\
Conselice C., 2014, ARAA, 52, 291 \\
daddi E., et al., 2007, ApJ, 670, 156 \\
Davies J., et al., 2014, MNRAS, 438, 1922 \\
Davies J., et al., 2017, PASP, 129, 4102 \\
Davies L. J. M., et al., 2015, MNRAS, 452, 616 \\
Davies L. J. M., et al., 2016a, MNRAS, 455, 4013 \\
Davies L. J. M., et al., 2016b, MNRAS, 461, 458 \\
De Lucia G., Weinmann S., Poggianti B., Aragon-Salamanca A. and Zaritsky D., 2012, MNRAS, 423, 1277 \\
De Vis P., et al., 2017a, MNRAS, 464, 4680 \\
De Vis et al., 2017b, MNRAS, 471, 1743 \\
De Vis P., et al., 2018, MNRAS, submitted \\
Dressler A., 1980, ApJ, 236, 351 \\
Dressler A., Thompson I. B., and Schechtman S. A., 1985, ApJ, 288, 481 \\
Dunne L., et al., 2011, MNRAS, 417, 1510 \\
Edmunds M., 1990, MNRAS, 246, 678 \\
Edmunds M. and Eales S., 1998, MNRAS, 299, L29 \\
Eales S., et al., 2017, MNRAS, 465, 3125 \\
Fumagalli M, Krumholz M., Prochaska J., Gavazzi G. and  Boselli A., 2009, ApJ, 697, 1811 \\
Gadotti, D., 2009, MNRAS, 393, 1531 \\
Gallazzi A. et al., 2009, ApJ, 690, 1883 \\
Gavazzi G., Boselli A., Pedotti P., Gallazzi A. and Carrasco L. 2002, AA, 396, 449 \\
Genzel R., et al., 2015, ApJ, 800, 20 \\
Giovanelli R. and Haynes M., 1985, ApJ, 292, 404 \\
Gomez P., et al., 2003, ApJ, 584, 210 \\
Gordon K. D., Clayton G. C., Witt, A. N., and Misselt K. A., 2000, ApJ, 533, 236 \\
Goudfrooij P. and de Jong T., AA, 298, 784 \\
Griffin M., et al., 2010, AA, 518, 3 \\
Jones A., Kohler M., Ysard N., Bocchio M., Verstraete L., 2017, AA, 602, 46 \\
Kauffmann G., et al., 2004, MNRAS, 353, 713 \\
Kennicutt R., 1983, AJ, 88, 483 \\
Kennicutt R., Bothun G. and Schommer R., 1984, AJ, 89, 1279 \\
Kodama T., Smail I., Nakata F., Okamura S. and Bower R., 2001, ApJ, 526, 9 \\
Kuutma T., Tamm A., Tempel E., 2017, AA, 600, 6 \\
Lebouteiller V., et al., 2017, AA, 602, 45 \\
Lofthouse E., Kaviraj S., Conselice C., Mortlock A. and Hartley W., 2017, MNRAS, 465, 2895 \\
Malavasi N., et al., 2017, MNRAS, 465, 3817 \\
McConnachie A., 2012, AJ, 144, 4 \\
Mclaughlin D., 1999, ApJ, 512, 9 \\
Moore B., Katz N., Lake G., Dressler A., Oemler A., 1996, Nature, 379, 613 \\
Moss C. and Whittle M., 1993, ApJ, 407, 17 \\
Muldrew S. et al., 2012, MNRAS, 419, 2670 \\
Nersesian A., et al., 2018, in press \\
Oemler A., 1974, ApJ, 194, 1 \\
Ostriker J. and  Tremaine S., 1975, ApJ, 202, 113\\
Park C., Gott R. and Choi Y., 2007, ApJ, 658, 898 \\
Park C., Gott J. Richard III, Choi Y., 2008, ApJ, 674, 784 \\
Peng Y., et al., 2010. ApJ, 721, 193 \\
Poglitsch et al., 2010, AA, 518, 2 \\
Poglitsch et al., 2017, ApJ, 844, 48 \\
Riess A. G., Macri L. M., Hoffmann, S. L., et al. 2016, ApJ, 826, 56 \\
Robotham A., et al., 2014, MNRAS, 444, 3986 \\
Rodriguez-Gomez V., et al., 2016, MNRAS, 458, 2371 \\
Rodriguez-Gomez V., et al., 2017, MNRAS, 467, 3083 \\
Roehlly, Y., Burgarella, D., Buat, V., et al. 2014, in Astronomical Society of the
Pacific Conference Series, Vol. 485, Astronomical Data Analysis Software
and Systems XXIII, ed. N. Manset \& P. Forshay, 347 \\
Saales L., et al., 2012, MNRAS, 423, 1544 \\
Schiminovich D., et al., 2007, ApJS, 173, 315 \\
Scoville N., et al., 2016, ApJ, 820, 83 \\
Scudder J., Ellison S., Torrey P., Patton D. and Mendel, J., 2012, MNRAS, 426, 549 \\
Smethurst R., et al., 2015, MNRAS, 450, 435 \\
Smith M., et al., 2012, ApJ, 748, 123 \\
Springel V., et al., 2005, Nature, 435, 629 \\
Strateva I., 2001, AJ, 122, 1861 \\
Tomczak A., et al., 2014, ApJ, 783, 85 \\
van der Kruit P. and Freeman K., 2011, ARAA, 49, 301 \\
Viaene S., et al., 2016, AA, 586, 13 \\
Visvanathan N. and Sandage A., 1977, ApJ, 216, 214 \\ 
Wang Y., et al., 2018, arXiv:1809.05244 \\
Weinmann S., van den Bosch F., Yang X. and Mo H., 2006, MNRAS, 366, 2 \\
Weinmann S., et al., 2009, MNRAS, 394, 1213 \\
Wel A., Bell E., Holden B., Skibba R., Rix H-W., 2010, ApJ, 714, 1279 \\
Whitaker K., 2015, ApJ, 811, 12 \\
White S. and Rees M., 1978, MNRAS, 183, 341 \\
Willett K., et al., 2015, MNRAS, 449, 820 \\

\end{document}